\documentclass[aps,superscriptaddress,preprint,pra,epsfig,showpacs,amsmath,amsfonts,amssymb,floatfix]{revtex4-1}
\usepackage{amsmath,amsfonts,amssymb,color}
\usepackage{amsthm}
\usepackage{leftidx}
\usepackage{graphicx}
\usepackage{xcolor}
\usepackage{dcolumn}
\usepackage{bm}
\usepackage{epstopdf}
\usepackage{epsfig}
\usepackage{environ}
\usepackage{pdfcomment}

\usepackage{float}
\usepackage[T1]{fontenc}
\usepackage[latin9]{inputenc}
\usepackage{setspace}
\usepackage{esint}

\begin{document}

\title{Generating many Majorana corner modes and multiple phase transitions in Floquet second-order topological superconductors}

\author{Longwen Zhou}
\email{zhoulw13@u.nus.edu}
\affiliation{%
	College of Physics and Optoelectronic Engineering, Ocean University of China, Qingdao, China 266100
}

\date{\today}

\begin{abstract}
A $d$-dimensional, $n$th-order topological insulator or superconductor
has localized eigenmodes at its $(d-n)$-dimensional boundaries ($n\leq d$).
In this work, we apply periodic driving fields to two-dimensional
superconductors, and obtain a wide variety of Floquet second-order
topological superconducting (SOTSC) phases with many Majorana corner
modes at both zero and $\pi$ quasienergies. Two distinct Floquet
SOTSC phases are found to be separated by three possible kinds of
transformations, i.e., a topological phase transition due to the closing/reopening
of a bulk spectral gap, a topological phase transition due to the
closing/reopening of an edge spectral gap, or an entirely different
phase in which the bulk spectrum is gapless. Thanks to the strong
interplay between driving and intrinsic energy scales of the system,
all the found phases and transitions are highly controllable via tuning
a single hopping parameter of the system. Our discovery not only enriches
the possible forms of Floquet SOTSC phases, but also offers an efficient scheme
to generate many coexisting Majorana zero and $\pi$ corner modes
that may find applications in Floquet quantum computation.
\end{abstract}

\pacs{}
\keywords{}
\maketitle

Floquet topological phases appear in time-periodic driven systems
with nontrivial topological properties (see Refs.~\cite{FTPRev1,FTPRev2,FTPRev3,FTPRev4}
for reviews). Their characteristic features include symmetry classifications
unique to nonequilibrium states \cite{FTPCls1,FTPCls2,FTPCls3,FTPCls4},
anomalous bulk-boundary correspondence with no static analogy \cite{AFTP1,AFTP2,AFTP3,AFTP4},
and substantial numbers of edge states induced by driving fields \cite{ManyFEM1,ManyFEM2,ManyFEM3,ManyFEM4,ManyFEM5}.
The observation of Floquet topological matter in various physical
settings including solids \cite{FTPExpSlid1,FTPExpSlid2,FTPExpSlid3,FTPExpSlid4},
cold atoms \cite{FTPExpCdAt1,FTPExpCdAt2,FTPExpCdAt3,FTPExpCdAt4}
and optical systems \cite{FTPExpOpt1,FTPExpOpt2,FTPExpOpt3,FTPExpOpt4}
further promotes their applications in ultrafast electronics \cite{FTPApp1}
and topological quantum computing \cite{FTPApp2}.

In recent years, the approach of Floquet engineering has been applied
to generate and control higher-order topological phases \cite{HOTP1,HOTP2,HOTP3,HOTP4,HOTP5,HOTP6,HOTP7,HOTP8,HOTP9}.
A gapped topological phase of order $n$ in $d$-spatial dimensions
($d\geq n\geq1$) usually holds topological edge states along its $(d-n)$-dimensional
boundaries (see Refs.~\cite{HOTPRev1,HOTPRev2,HOTPRev3,HOTPRev4} for reviews).
For example, a two-dimensional (2D), second-order topological insulator
has localized eigenstates at the corners of the lattice on which it
is defined. The conventional topological insulator can thus be viewed
as a first-order topological phase. Besides insulating \cite{FHOTP1,FHOTP2,FHOTP3,FHOTP4,FHOTP5,FHOTP6,FHOTP7,FHOTP8,FHOTP9,FHOTP10,FHOTP11,FHOTP12,FHOTP13,FHOTP14,FHOTP15,FHOTP16,FHOTP17,FHOTP18,FHOTP19,FHOTP20,FHOTP21,FHOTP22}
and semi-metallic \cite{FHOTSM1,FHOTSM2,FHOTSM3,FHOTSM4,FHOTSM5}
setups, periodic driving fields have also been applied to engineer
Floquet SOTSC phases \cite{FHOTSC1,FHOTSC2,FHOTSC3,FHOTSC4,FHOTSC5,FHOTSC6,FHOTSC7,FHOTSC8,FHOTSC9}.
The latter could possess two types of symmetry-protected Majorana
corner states with zero and $\pi$ quasienergies, which might be adopted
in topological quantum computation as substitutions of the more conventional
Majorana edge modes. Till now, the studies of Floquet second-order
topological superconductors focus on either models with unconventional
$d$-wave pairings and up to eight Floquet bands \cite{FHOTSC7,FHOTSC8,FHOTSC9},
or applications in Floquet quantum computation utilizing at most four
Majorana corner modes at both zero and $\pi$ quasienergies \cite{FHOTSC2,FHOTSC3}.
It remains unclear whether we could obtain as many as possible the
Majorana zero and $\pi$ corner modes together in Floquet SOTSC phases
following simple driving protocols, similar to what has been achieved
for Floquet second-order topological insulators \cite{FHOTP1}. Moreover,
phase transitions in higher-order topological matter could happen
by closing either a gap between the bulk bands (type-I) or a gap between
the edge bands (type-II)~\cite{HOTPT1,HOTPT2,HOTPT3,HOTPT4,HOTPT5,HOTPT6,HOTPT7}.
Following which type of topological phase transition could Floquet
Majorana corner modes emerge deserves to be further clarified.

In this work, we couple a $p+ip$ superconductor in two dimensions
to time-periodic driving fields, and obtain rich Floquet SOTSC phases
with many normal and anomalous Majorana corner modes at zero and $\pi$
quasienergies, respectively. In Sec.~\ref{sec:Mod}, we introduce
our model and its driving protocol, discuss the symmetries of the
model, and explain how to find its spectrum and eigenmodes under different
boundary conditions. In Sec.~\ref{sec:Res}, we explore the topological
phases and phase transitions in our system with gradually increased
generality. Floquet SOTSC phases with different numbers of Majorana
zero and $\pi$ corner modes are obtained, and they are found to be
separated by the closing/reopening of bulk spectral gaps, edge band
gaps or even emerging new phases with gapless Floquet spectra. In
Sec.~\ref{sec:Sum}, we summarize our results and discuss potential
future directions. Throughout this work, we set the Floquet driving
period $T=1$ and the Planck constant $\hbar=1$. Other system parameters
have been properly scaled and set in dimensionless units.
\section{Model}\label{sec:Mod}

\begin{figure}
	\begin{centering}
		\includegraphics[width=10.5 cm]{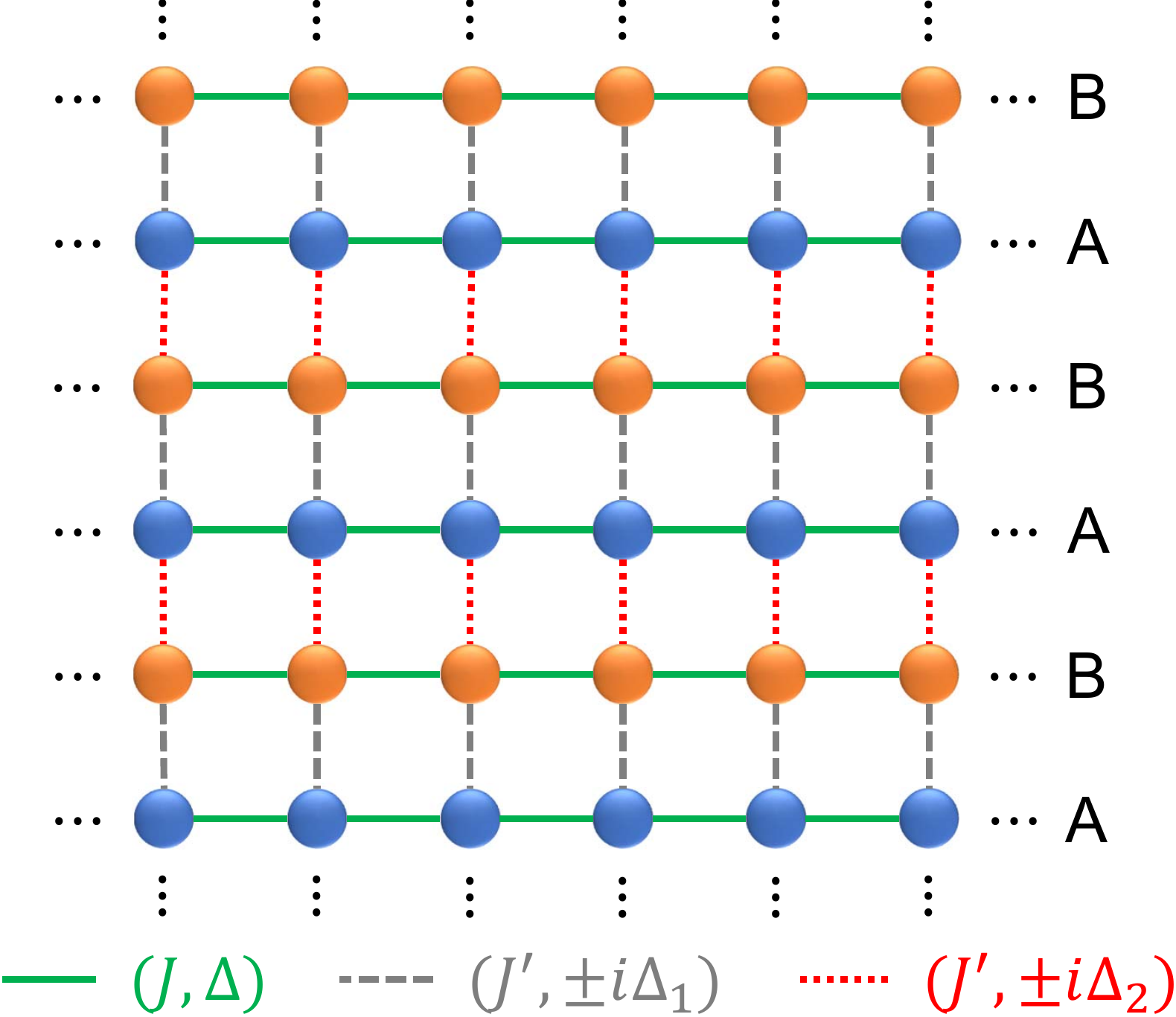}
		\par\end{centering}
	\caption{An illustration of the 2D lattice model. The chains A (blue balls) and B (orange balls) are related to the operators ${\hat a}^{(\dagger)}$ and ${\hat b}^{(\dagger)}$ in ${\hat H}$. A uniform chemical potential $\mu$ is applied to each site. In the Floquet model, the superconducting pairing amplitude $\Delta$ is replaced by $\Delta(t)=\sum_\ell\delta(t-\ell)$.}\label{fig:Sketch}
\end{figure}

\begin{figure}
	\begin{centering}
		\includegraphics[width=10.5 cm]{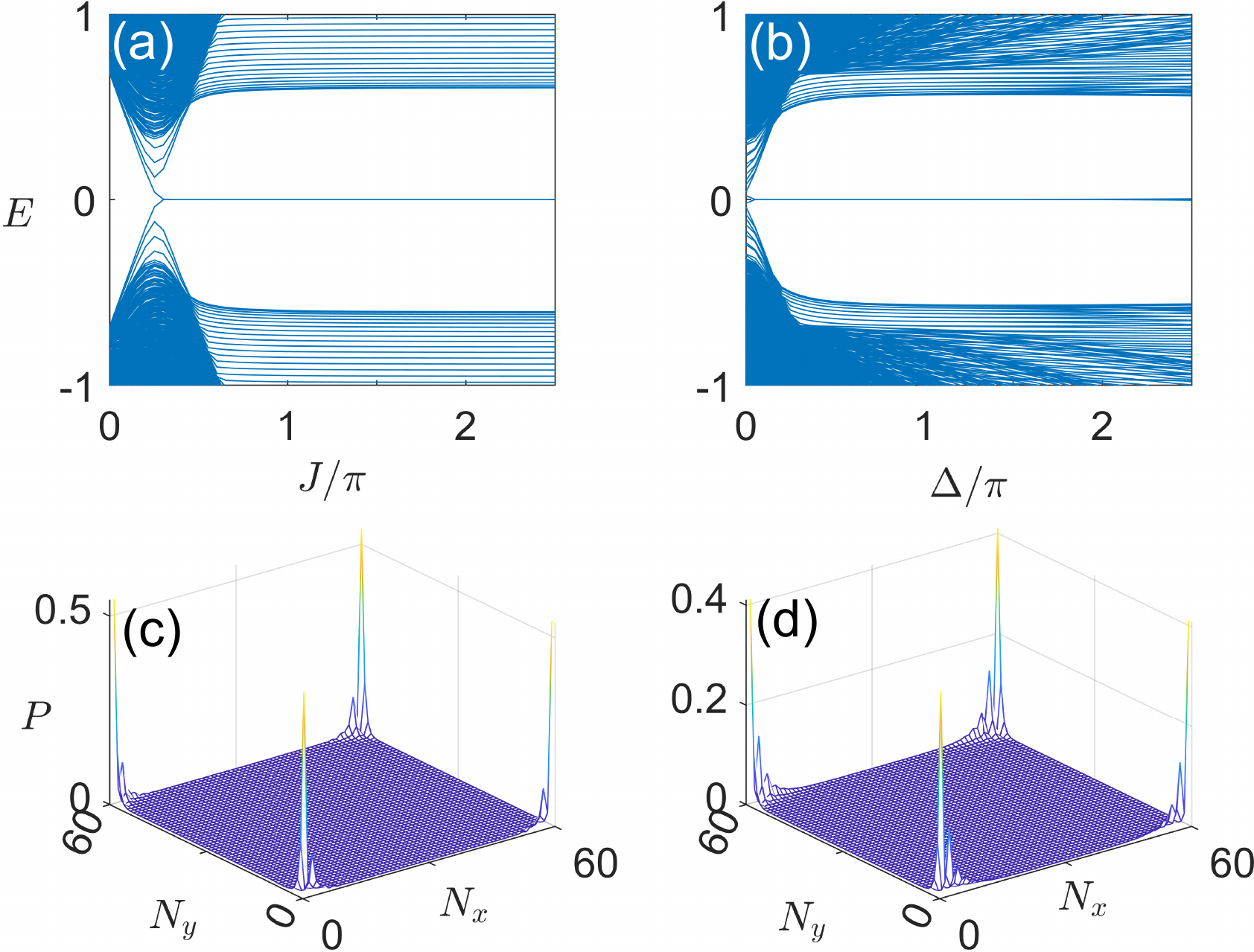}
		\par\end{centering}
	\caption{Typical spectra and Majorana corner modes of the static SOTSC under open boundary conditions along $x$ and $y$ directions. (a) and (b) show the energy spectra $E$ of ${\hat H}$ versus $J$ and $\Delta$, with other system parameters chosen to be $(\mu,J',\Delta_1,\Delta_2)=(0.25\pi,0.05\pi,0.2\pi,0.4\pi)$. (c) and (d) show the probability distributions of the four Majorana corner modes with $E=0$ in (a) and (b) at $J=1.5\pi$ and $\Delta=1.5\pi$, respectively. The lattice sizes are set as $N_x=N_y=60$ for all panels.\label{fig:SOTSC}}
\end{figure}

We start with a lattice model that describes a 2D $p_{x}+ip_{y}$ superconductor.
It may be viewed as a simplified version of the model introduced in the Eq.~(1) of Ref.~\cite{FHOTSC2}.
A schematic diagram of our model is shown in Fig.~\ref{fig:Sketch}, whose static Hamiltonian takes the form

\begin{equation}
	\hat{H}=\hat{H}_{x}+\hat{H}_{y},\label{eq:H}
\end{equation}
where

\begin{alignat}{1}
	\hat{H}_{x}= & \frac{1}{2}\sum_{m,n}\left[\mu\left(\hat{a}_{m,n}^{\dagger}\hat{a}_{m,n}-\frac{1}{2}\right)+J\hat{a}_{m,n}^{\dagger}\hat{a}_{m+1,n}+\Delta\hat{a}_{m,n}\hat{a}_{m+1,n}\right]\label{eq:HX}\\
	+ & \frac{1}{2}\sum_{m,n}\left[\mu\left(\hat{b}_{m,n}^{\dagger}\hat{b}_{m,n}-\frac{1}{2}\right)+J\hat{b}_{m,n}^{\dagger}\hat{b}_{m+1,n}+\Delta\hat{b}_{m,n}\hat{b}_{m+1,n}\right]+{\rm H.c.},\nonumber 
\end{alignat}
and

\begin{equation}
	\hat{H}_{y}= \sum_{m,n}J'\left(\hat{a}_{m,n}^{\dagger}\hat{b}_{m,n}+\hat{b}_{m,n}^{\dagger}\hat{a}_{m,n+1}\right)
	-\sum_{m,n}i\left(\Delta_{1}\hat{a}_{m,n}\hat{b}_{m,n}+\Delta_{2}\hat{b}_{m,n}\hat{a}_{m,n+1}\right)+{\rm H.c.}.\label{eq:HY}
\end{equation}
Here $\hat{a}_{m,n}^{\dagger}$ ($\hat{a}_{m,n}$) and $\hat{b}_{m,n}^{\dagger}$
($\hat{b}_{m,n}$) are the creation (annihilation) operators of fermions
on the sublattices A and B of a unit cell at the location ${\bf r}=(m,n)$ of the 2D lattice.
The sublattice structure is originated from the dimerized superconducting
pairing amplitudes ($\Delta_{1}\neq\Delta_{2}$) along the $y$-direction
of the lattice. $J$ and $J'$ denote the hopping amplitudes along
the $x$- and $y$-directions. $\mu$ is the chemical potential and
$\Delta$ is the pairing amplitude along the $x$-direction.
By solving the eigenvalue equation ${\hat H}|\psi\rangle=E|\psi\rangle$, one can obtain the spectrum and Majorana corner states of the static system, with typical examples shown in Fig.~\ref{fig:SOTSC}. We observe that there are two possible phases, with one of them being topologically trivial and the other one possessing four zero-energy Majorana eigenmodes at the four corners of the lattice. For the parameter domains considered in Fig.~\ref{fig:SOTSC}, these Majorana corner modes emerge once $|J|>|\mu|$. Similar results are observed in the previous study of a slightly more complicated model \cite{FHOTSC2}, where the hopping amplitude $J'$ and chemical potential $\mu$ are also dimerized in space.
In Ref. \cite{FHOTSC2}, it was shown that with a harmonic (sinusoidal) driving field added
to the chemical potential $\mu$, the system could become a Floquet
second-order topological superconductor with four Majorana corner
modes at both the zero and $\pi$ quasienergies. 
Note that the model
used in Ref.~\cite{FHOTSC2} further assumes the lattice dimerization
in the hopping amplitude $J'$ and chemical potential $\mu$. In our
model above, the hopping and chemical potential terms are instead uniform along
the two spatial dimensions. Besides, the Floquet driving field in Ref.~\cite{FHOTSC2} was added
to the dimerized chemical potential there. In our case, the driving field will be coupled to the superconducting pairing amplitude as discussed below. 
The possibility of generating more Floquet Majorana zero/$\pi$ corner modes and inducing multiple
topological transitions by periodic driving fields were also not considered in Ref.~\cite{FHOTSC2},
as the sinusoidal modulation applied there could not achieve such goals.

In this work, we consider a different driving protocol by applying
time-periodic kicks to the superconducting pairing amplitude $\Delta$
along the $x$-direction, i.e., by setting $\Delta\rightarrow\Delta(t)$
in Eq. (\ref{eq:HX}) with $\Delta(t)=\Delta\sum_{\ell\in\mathbb{Z}}\delta(t-\ell)$.
In one-dimensional Floquet topological superconductors, this type of driving and its physical relevance has been considered in \cite{NHFTSC1}. 
The Floquet operator of the system, which describes its evolution
over a complete driving period (e.g., from $t=0^{-}\rightarrow1^{-}$)
then takes the form

\begin{equation}
	\hat{U}=\hat{U}_{2}\hat{U}_{1},\label{eq:U}
\end{equation}
where

\begin{equation}
	\hat{U}_{1}=e^{-\frac{i}{2}\sum_{m,n}\Delta\left(\hat{a}_{m,n}\hat{a}_{m+1,n}+\hat{b}_{m,n}\hat{b}_{m+1,n}+{\rm H.c.}\right)},\label{eq:U1}
\end{equation}
and

\begin{equation}
	\hat{U}_{2}=e^{-i\hat{H}_{y}-\frac{i}{2}\sum_{m,n}\left[\mu\left(\hat{a}_{m,n}^{\dagger}\hat{a}_{m,n}+\hat{b}_{m,n}^{\dagger}\hat{b}_{m,n}-1\right)+J\left(\hat{a}_{m,n}^{\dagger}\hat{a}_{m+1,n}+\hat{b}_{m,n}^{\dagger}\hat{b}_{m+1,n}\right)+{\rm H.c.}\right]}.\label{eq:U2}
\end{equation}
The Floquet spectrum and eigenstates of $\hat{U}$ can be obtained
from the eigenvalue equation $\hat{U}|\psi\rangle=e^{-iE}|\psi\rangle$,
whose solution $|\psi\rangle$ describes a Floquet state with the
quasienergy $E\in[-\pi,\pi)$ (defined modulus $2\pi$). Under open
boundary conditions (OBCs) along both the $x$- and $y$-directions,
we refer to a corner-localized eigenmode of $\hat{U}$ with the quasienergy
$E=0$ ($E=\pm\pi$) as a Floquet Majorana zero ($\pi$) corner mode.
In Sec.~\ref{sec:Res}, we will show that many such Floquet Majorana
corner modes could emerge in our system. Moreover, their numbers can
be tuned by changing the system parameter across different types of
topological phase transitions.

Under periodic boundary conditions (PBCs) along both the $x$- and
$y$-directions, we can apply Fourier transformations to the creation
and annihilation operators, i.e.,

\begin{equation}
	\hat{a}_{{\bf r}}=\frac{1}{\sqrt{S}}\sum_{{\bf k}}e^{i{\bf k}\cdot{\bf r}}\hat{a}_{{\bf k}},\qquad\hat{a}_{{\bf r}}^{\dagger}=\frac{1}{\sqrt{S}}\sum_{{\bf k}}e^{-i{\bf k}\cdot{\bf r}}\hat{a}_{{\bf k}}^{\dagger},\label{eq:ak}
\end{equation}
\begin{equation}
	\hat{b}_{{\bf r}}=\frac{1}{\sqrt{S}}\sum_{{\bf k}}e^{i{\bf k}\cdot{\bf r}}\hat{b}_{{\bf k}},\qquad\hat{b}_{{\bf r}}^{\dagger}=\frac{1}{\sqrt{S}}\sum_{{\bf k}}e^{-i{\bf k}\cdot{\bf r}}\hat{b}_{{\bf k}}^{\dagger}.\label{eq:bk}
\end{equation}
Here $S=N_{x}N_{y}$ denotes the total number of unit cells, with
$N_{x}$ ($N_{y}$) being the number of cells along the $x$ ($y$)
direction of the lattice. The unit cell index is ${\bf r}=(m,n)$,
where $m=1,2,...,N_{x}-1,N_{x}$ and $n=1,2,...,N_{y}-1,N_{y}$. We
thus identify $\hat{a}_{{\bf r}}=\hat{a}_{m,n}$ and $\hat{b}_{{\bf r}}=\hat{b}_{m,n}$
for the second-quantized operators. The 2D quasimomentum is ${\bf k}=(k_{x},k_{y})$,
with $k_{x}=-\pi,-\pi+\frac{2\pi\cdot1}{N_{x}},...,-\pi+\frac{2\pi\cdot(N_{x}-2)}{N_{x}},-\pi+\frac{2\pi\cdot(N_{x}-1)}{N_{x}}$
($k_{y}=-\pi,-\pi+\frac{2\pi\cdot1}{N_{y}},...,-\pi+\frac{2\pi\cdot(N_{y}-2)}{N_{y}},-\pi+\frac{2\pi\cdot(N_{y}-1)}{N_{y}}$
) being the quasimomentum along the $x$ ($y$) direction. Therefore
we also identify $\hat{a}_{{\bf k}}=\hat{a}_{k_{x},k_{y}}$ and $\hat{b}_{{\bf k}}=\hat{b}_{k_{x},k_{y}}$.
After the Fourier transformation, we find the components $\hat{H}_{x}(t)$
and $\hat{H}_{y}$ of the driven Hamiltonian $\hat{H}(t)=\hat{H}_{x}(t)+\hat{H}_{y}$ as

\begin{equation}
	\hat{H}_{x}(t)=\frac{1}{2}\sum_{{\bf k}}\hat{\Psi}_{{\bf k}}^{\dagger}H_{x}({\bf k},t)\hat{\Psi}_{{\bf k}},\qquad\hat{H}_{y}=\frac{1}{2}\sum_{{\bf k}}\hat{\Psi}_{{\bf k}}^{\dagger}H_{y}({\bf k})\hat{\Psi}_{{\bf k}}.\label{eq:Hk}
\end{equation}
Here $\hat{\Psi}_{{\bf k}}^{\dagger}=(\hat{a}_{{\bf k}}^{\dagger},\hat{a}_{-{\bf k}},\hat{b}_{{\bf k}}^{\dagger},\hat{b}_{-{\bf k}})$
is the creation operator in the Nambu basis. The Hamiltonian matrices
$H_{x}({\bf k},t)$ and $H_{y}({\bf k})$ in the Nambu spinor representation are given by

\begin{equation}
	H_{x}({\bf k},t)=\sigma_{0}\otimes\left[\Delta(t)\sin k_{x}\sigma_{y}+(\mu+J\cos k_{x})\sigma_{z}\right],\label{eq:HXk}
\end{equation}
\begin{equation}
	H_{y}({\bf k})= J'\left[(1+\cos k_{y})\sigma_{x}+\sin k_{y}\sigma_{y}\right]\otimes\sigma_{z}
	+\left[\Delta_{2}\sin k_{y}\sigma_{x}+(\Delta_{1}-\Delta_{2}\cos k_{y})\sigma_{y}\right]\otimes\sigma_{x},\label{eq:HYk}
\end{equation}
where $\sigma_{x,y,z}$ are Pauli matrices and $\sigma_{0}$ is the
two by two identity matrix. In the momentum space, the Floquet operator now takes the form

\begin{equation}
	U({\bf k})=U_{2}({\bf k})U_{1}({\bf k})=e^{-iH_{2}({\bf k})}e^{-iH_{1}({\bf k})},\label{eq:Uk}
\end{equation}
where

\begin{equation}
	H_{1}({\bf k})=\Delta(t)\sin k_{x}\sigma_{0}\otimes\sigma_{y},\label{eq:H1k}
\end{equation}
\begin{alignat}{1}
	H_{2}({\bf k})& = (\mu+J\cos k_{x})\sigma_{0}\otimes\sigma_{z}+J'\left[(1+\cos k_{y})\sigma_{x}+\sin k_{y}\sigma_{y}\right]\otimes\sigma_{z}\nonumber \\
	& + \left[\Delta_{2}\sin k_{y}\sigma_{x}+(\Delta_{1}-\Delta_{2}\cos k_{y})\sigma_{y}\right]\otimes\sigma_{x}.\label{eq:H2k}
\end{alignat}
The Floquet spectrum and eigenstates of the system in ${\bf k}$-space
are then obtained by solving the $4\times4$ eigenvalue equation $U({\bf k})|\psi_{j}({\bf k})\rangle=e^{-iE_{j}({\bf k})}|\psi_{j}({\bf k})\rangle$,
yielding four quasienergy bands $E_{j}({\bf k})$ for $j=1,2,3,4$.
When there exists $j$ and $l$ such that $E_{j}({\bf k})=E_{l}({\bf k})=0$
or $E_{j}({\bf k})=E_{l}({\bf k})=\pm\pi$, the bulk Floquet spectral
gap may close at the center or boundary of the first quasienergy Brillouin
zone $E\in[-\pi,\pi)$. In Sec.~\ref{sec:Res}, we will explore the
Floquet spectra and gap closing conditions of our system for cases
with gradually increased complexity. We will further relate the gap
closing/reopening transitions of the bulk Floquet spectra with topological
phase transitions accompanied by the adjustment of Floquet corner
modes under OBCs.

The topological properties of the system is closely related to the
symmetries of $U({\bf k})$. Upon unitary transformations, we can
express the Floquet operator in a pair of symmetric time frames \cite{STM1,STM2,STM3,STM4} as

\begin{equation}
	{\cal U}_{1}({\bf k})=U_{1}^{1/2}({\bf k})U_{2}({\bf k})U_{1}^{1/2}({\bf k}),\qquad {\cal U}_{2}({\bf k})=U_{2}^{1/2}({\bf k})U_{1}({\bf k})U_{2}^{1/2}({\bf k}).\label{eq:UTF}
\end{equation}
When the hopping amplitude $J'=0$, it is not hard to verify that
both ${\cal U}_{1}({\bf k})$ and ${\cal U}_{2}({\bf k})$ possess
the chiral symmetry $\Gamma=\sigma_{z}\otimes\sigma_{x}$ in the sense
that $\Gamma{\cal U}_{\alpha}({\bf k})\Gamma={\cal U}_{\alpha}^{-1}({\bf k})$
for $\alpha=1,2$. Since unitary transformations do not change the
quasienergy spectrum, the chiral symmetry $\Gamma$ enforces the degeneracy
of the eigenmodes of $U({\bf k})$ when their quasienergies are equal
to zero or $\pm\pi$ under the condition $J'=0$. This will be the
situation in the first three cases explored in Sec.~\ref{sec:Res}.
When $J'\neq0$, the chiral symmetry $\Gamma$ is broken. But ${\cal U}_{1}({\bf k})$
and ${\cal U}_{2}({\bf k})$ still possess the particle-hole symmetry
${\cal C}=\sigma_{0}\otimes\sigma_{x}{\cal K}$, enforcing the Floquet
spectrum of $U({\bf k})$ to be symmetric with respect to the quasienergy
$E=0$. With the particle-hole symmetry only, we may obtain at most
four Floquet Majorana corner modes at the quasienergies zero and $\pm\pi$
when a rectangle geometry and the OBCs along both $x$ and $y$ directions
are taken for the system. This will be the case encountered in Sec.~\ref{subsec:C4}.

In higher-order topological phases, the change of topological corner
modes can be induced not only by closing/reopening a bulk spectrum
gap, but also by closing/reopening a gap between different edge bands.
The former (latter) is usually called a type-I (type-II) topological
phase transition \cite{HOTPT1,HOTPT2,HOTPT3,HOTPT4,HOTPT5,HOTPT6,HOTPT7}. To unveil the possibility of generating
these two types of phase transitions in Floquet systems, we will also
consider the solution of the system under the PBC (OBC) along $x$
($y$) direction of the lattice, or vice versa. In the former case,
we can apply the Fourier transformation only along the $x$-direction, i.e.,

\begin{equation}
	\hat{a}_{m,n}=\frac{1}{\sqrt{N_{x}}}\sum_{k_{x}}e^{ik_{x}m}\hat{a}_{k_{x},n},\qquad\hat{a}_{m,n}^{\dagger}=\frac{1}{\sqrt{N_{x}}}\sum_{k_{x}}e^{-ik_{x}m}\hat{a}_{k_{x},n}^{\dagger},\label{eq:ax}
\end{equation}
\begin{equation}
	\hat{b}_{m,n}=\frac{1}{\sqrt{N_{x}}}\sum_{k_{x}}e^{ik_{x}m}\hat{b}_{k_{x},n},\qquad\hat{b}_{m,n}^{\dagger}=\frac{1}{\sqrt{N_{x}}}\sum_{k_{x}}e^{-ik_{x}m}\hat{b}_{k_{x},n}^{\dagger}.\label{eq:bx}
\end{equation}
The Floquet operator under the PBC (OBC) along $x$ ($y$) direction then takes the form

\begin{equation}
	\hat{U}(k_{x})=\hat{U}_{2}(k_{x})\hat{U}_{1}(k_{x}),\label{eq:Ukx}
\end{equation}
where

\begin{equation}
	\hat{U}_{1}=e^{-\frac{i}{2}\sum_{m,n}\Delta\left(\hat{a}_{k_{x},n}\hat{a}_{-k_{x},n}e^{-ik_{x}}+\hat{b}_{k_{x},n}\hat{b}_{-k_{x},n}e^{-ik_{x}}+{\rm H.c.}\right)},\label{eq:U1kx}
\end{equation}
\begin{equation}
	\hat{U}_{2}=e^{-i\hat{H}_{y}(k_{x})-\frac{i}{2}\sum_{m,n}\left[\mu\left(\hat{a}_{k_{x},n}^{\dagger}\hat{a}_{k_{x},n}+\hat{b}_{k_{x},n}^{\dagger}\hat{b}_{k_{x},n}-1\right)+Je^{ik_{x}}\left(\hat{a}_{k_{x},n}^{\dagger}\hat{a}_{k_{x},n}+\hat{b}_{k_{x},n}^{\dagger}\hat{b}_{k_{x},n}\right)+{\rm H.c.}\right]},\label{eq:U2kx}
\end{equation}
and

\begin{equation}
	\hat{H}_{y}(k_{x})=\sum_{k_{x},n}\left[J'\left(\hat{a}_{k_{x},n}^{\dagger}\hat{b}_{k_{x},n}+\hat{b}_{k_{x},n}^{\dagger}\hat{a}_{k_{x},n+1}\right)
	-i\left(\Delta_{1}\hat{a}_{k_{x},n}\hat{b}_{-k_{x},n}+\Delta_{2}\hat{b}_{k_{x},n}\hat{a}_{-k_{x},n+1}\right)\right]+{\rm H.c.}.\label{eq:HYkx}
\end{equation}
The Floquet spectrum and eigenstates are then obtained by diagonalizing
$\hat{U}(k_{x})$ at all different quasimomenta $k_{x}\in[-\pi,\pi)$.
Similarly, under the PBC (OBC) along $y$ ($x$) direction of the
lattice, we can apply the Fourier transformation along the $y$-direction, i.e.,

\begin{equation}
	\hat{a}_{m,n}=\frac{1}{\sqrt{N_{y}}}\sum_{k_{y}}e^{ik_{y}n}\hat{a}_{m,k_{y}},\qquad\hat{a}_{m,n}^{\dagger}=\frac{1}{\sqrt{N_{y}}}\sum_{k_{y}}e^{-ik_{y}n}\hat{a}_{m,k_{y}}^{\dagger},\label{eq:ay}
\end{equation}
\begin{equation}
	\hat{b}_{m,n}=\frac{1}{\sqrt{N_{y}}}\sum_{k_{y}}e^{ik_{y}n}\hat{b}_{m,k_{y}},\qquad\hat{b}_{m,n}^{\dagger}=\frac{1}{\sqrt{N_{y}}}\sum_{k_{y}}e^{-ik_{y}n}\hat{b}_{m,k_{y}}^{\dagger}.\label{eq:by}
\end{equation}
The resulting Floquet operator takes the form

\begin{equation}
	\hat{U}(k_{y})=\hat{U}_{2}(k_{y})\hat{U}_{1}(k_{y}),\label{eq:Uky}
\end{equation}
where

\begin{equation}
	\hat{U}_{1}=e^{-\frac{i}{2}\sum_{m,k_{y}}\Delta\left(\hat{a}_{m,k_{y}}\hat{a}_{m+1,-k_{y}}+\hat{b}_{m,k_{y}}\hat{b}_{m+1,-k_{y}}+{\rm H.c.}\right)},\label{eq:U1ky}
\end{equation}
\begin{equation}
	\hat{U}_{2}=e^{-i\hat{H}_{y}(k_{y})-\frac{i}{2}\sum_{m,k_{y}}\left[\mu\left(\hat{a}_{m,k_{y}}^{\dagger}\hat{a}_{m,k_{y}}+\hat{b}_{m,k_{y}}^{\dagger}\hat{b}_{m,k_{y}}-1\right)+J\left(\hat{a}_{m,k_{y}}^{\dagger}\hat{a}_{m+1,k_{y}}+\hat{b}_{m,k_{y}}^{\dagger}\hat{b}_{m+1,k_{y}}\right)+{\rm H.c.}\right]},\label{eq:U2ky}
\end{equation}
and

\begin{alignat}{1}
	\hat{H}_{y}(k_{y})= & \sum_{m,k_{y}}J'\left(\hat{a}_{m,k_{y}}^{\dagger}\hat{b}_{m,k_{y}}+\hat{b}_{m,k_{y}}^{\dagger}\hat{a}_{m,k_{y}}e^{ik_{y}}\right)\label{eq:HYky}\\
	- & \sum_{m,k_{y}}i\left(\Delta_{1}\hat{a}_{m,k_{y}}\hat{b}_{m,-k_{y}}+\Delta_{2}\hat{b}_{m,k_{y}}\hat{a}_{m,-k_{y}}e^{-ik_{y}}\right)+{\rm H.c.}.\nonumber 
\end{alignat}
The Floquet spectrum and eigenstates can now be found by diagonalizing
$\hat{U}(k_{y})$ at all different quasimomenta $k_{y}\in[-\pi,\pi)$.
The spectrum of $\hat{U}(k_{x})$ and $\hat{U}(k_{y})$ could provide
useful information for us to understand different types of phase transitions
in Floquet second-order topological superconductors, as will be shown
in the following section.

\section{Results}\label{sec:Res}

In this section, we study the emerging Floquet SOTSC phases in our
system with gradually increased complexity. We start with a minimal
model which yet possesses rich Floquet SOTSC phases, many Majorana
zero/$\pi$ corner modes and different classes of topological phase
transitions in Sec.~\ref{subsec:C1}. How will these intriguing states
and transitions be modified by the presence of finite chemical potential
$\mu$ and intracell superconducting pairing $\Delta_{1}$ are further
explored in Secs.~\ref{subsec:C2} and \ref{subsec:C3}. The most
general case with nonvanishing hopping amplitude $J'$ is finally
discussed in Sec.~\ref{subsec:C4}.

\subsection{Case 1: $\mu=J'=\Delta_1=0$}\label{subsec:C1}

\begin{figure}
	\begin{centering}
		\includegraphics[width=10.5 cm]{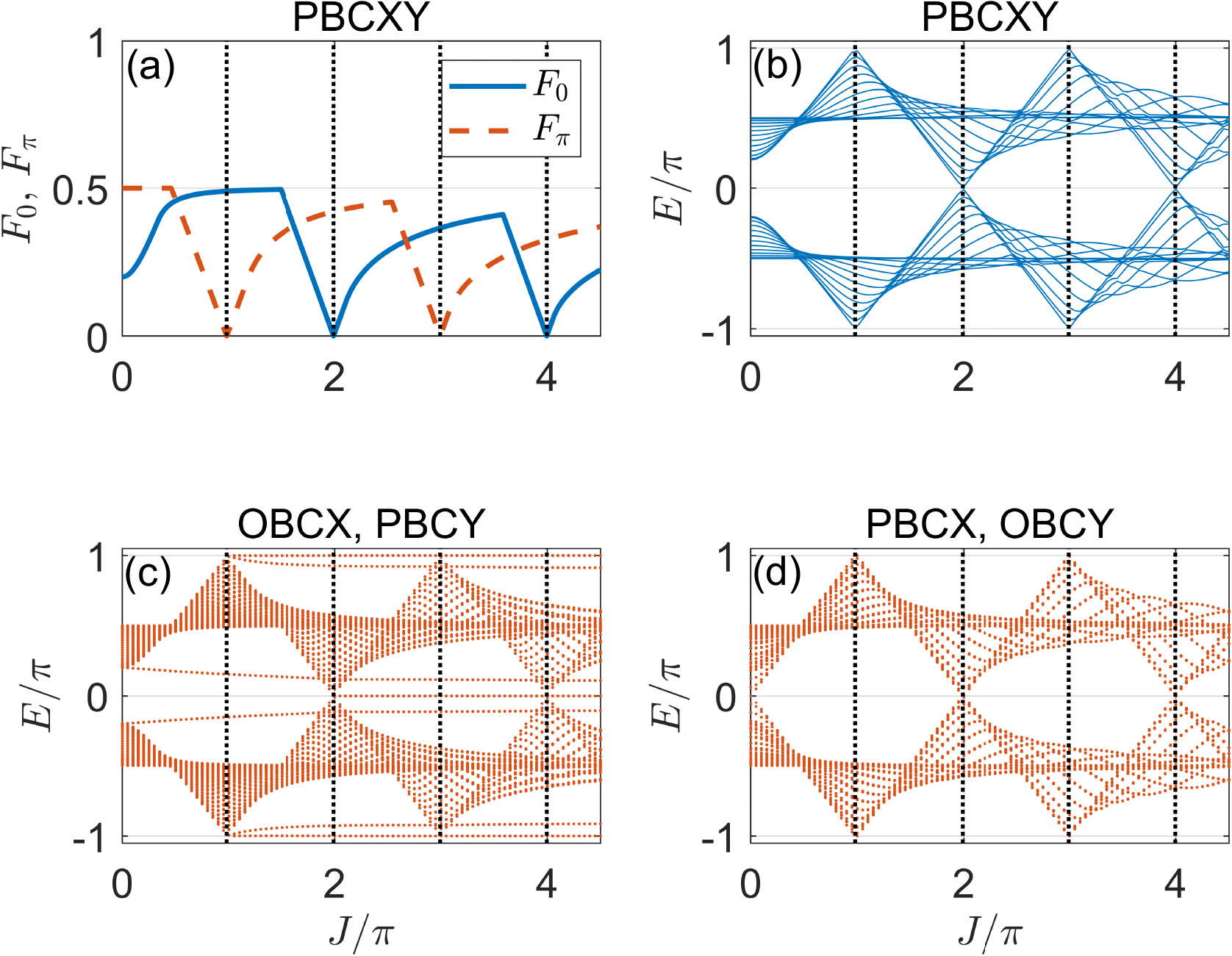}
		\par\end{centering}
	\caption{Gap functions and Floquet spectrum versus $J$ under different boundary
		conditions in the Case 1. (a) Gap functions under PBCXY. (b) Floquet spectrum
		under PBCXY. (c) Floquet spectrum under OBCX, PBCY. (d) Floquet spectrum
		under PBCX, OBCY. The crossing points between the vertical dotted
		lines and the horizontal axis show the bulk gap-closing points predicted
		by Eq.~(\ref{eq:C1PB}). Other system parameters are set as $(\Delta,\Delta_{2})=(0.5\pi,0.2\pi)$.
		The number of cells along $x$ and $y$ directions are $N_{x}=N_{y}=60$.
		\label{fig:C1GF}}
\end{figure}

We start with a simplest construction that allows us to have Floquet
SOTSC phases with Majorana corner modes in our system. Assuming the
chemical potential $\mu$, hopping amplitude $J'$ and intracell pairing
amplitude $\Delta_{1}$ along the $y$-direction to be zero, we arrive
at following Floquet operator in momentum space from Eqs. (\ref{eq:Uk})--(\ref{eq:H2k}), i.e.,

\begin{equation}
	U({\bf k})=e^{-i[J\cos k_{x}\sigma_{0}\otimes\sigma_{z}+\Delta_{2}(\sin k_{y}\sigma_{x}-\cos k_{y}\sigma_{y})\otimes\sigma_{x}]}
	e^{-i\Delta\sin k_{x}\sigma_{0}\otimes\sigma_{y}}.\label{eq:C1U}
\end{equation}
It is clear that the tensor product matrices $\sigma_{0}\otimes\sigma_{z}$,
$\sigma_{x}\otimes\sigma_{x}$, $\sigma_{y}\otimes\sigma_{x}$ and
$\sigma_{0}\otimes\sigma_{y}$ are anti-commute with one another.
We can thus apply the Taylor expansion to each exponential term in
Eq. (\ref{eq:C1U}). After recombining relevant terms in the expansion,
we find the quasienergy band dispersions

\begin{equation}
	E_{\pm}({\bf k})=\pm\arccos\left[\cos(\Delta\sin k_{x})\cos\left(\sqrt{J^{2}\cos^{2}k_{x}+\Delta_{2}^{2}}\right)\right].\label{eq:C1E}
\end{equation}
This gives us two quasienergy bands, with each of them being twofold
degenerate. Moreover, the dispersions $E_{\pm}({\bf k})$ are independent
of $k_{y}$, which means that all the Floquet bands of our system in
Case 1 are flat along $k_{y}$. Since the two pairs of bands are symmetric
with respect to the zero quasienergy, they could touch with each other
either at $E_{\pm}({\bf k})=0$ (center of the quasienergy Brillouin
zone) or at $E_{\pm}({\bf k})=\pm\pi$ (edge of the quasienergy Brillouin
zone). The Floquet spectrum could then become gapless if

\begin{equation}
	\cos(\Delta\sin k_{x})\cos\left(\sqrt{J^{2}\cos^{2}k_{x}+\Delta_{2}^{2}}\right)=\pm1,\label{eq:C1PB0}
\end{equation}
whose solution yields the phase boundary equation in parameter space

\begin{equation}
	\frac{p^{2}\pi^{2}}{\Delta^{2}}+\frac{q^{2}\pi^{2}-\Delta_{2}^{2}}{J^{2}}=1,\qquad p,q\in\mathbb{Z}.\label{eq:C1PB}
\end{equation}
That is, when the system parameters satisfy this equation, the bulk
quasienergy spectrum becomes gapless at either $E=0$ or $E=\pi$.
To investigate gap-closing transitions induced by the change of system
parameters, we introduce the following quasienergy gap functions

\begin{equation}
	F_{0}\equiv \frac{1}{\pi}\min_{{\bf k}\in{\rm BZ}}|E_{\pm}({\bf k})|,\qquad
	F_{\pi}\equiv \frac{1}{\pi}\min_{{\bf k}\in{\rm BZ}}||E_{\pm}({\bf k})|-\pi|.\label{eq:F0P}
\end{equation}
Here the $1/\pi$ in front is a scaling factor that restricts the
ranges of both $F_{0}$ and $F_{\pi}$ to $[0,1]$. It is clear that
$F_{0}=0$ ($F_{\pi}=0$) once the spectrum gap closes at the quasienergy
zero ($\pi$). In Fig.~\ref{fig:C1GF}(a), we show the gap functions
$F_{0}$ and $F_{\pi}$ versus the hopping amplitude $J$ for a typical
set of system parameters in Case 1. For our choice of system parameters,
we have $p=0$, and according to Eq. (\ref{eq:C1PB}) the phase transition
points appear at $J=\sqrt{q^{2}\pi^{2}-\Delta_{2}^{2}}$ for $q\in\mathbb{Z}$
assuming $q^{2}\pi^{2}\geq\Delta_{2}^{2}$. They are coincide with
the locations where $F_{0}$ or $F_{\pi}$ vanish in Fig.~\ref{fig:C1GF}(a).
Moreover, we observe a series of gap closing transitions with the
increase of $J$, and more such transitions are expected to happen
at larger values of $J$, which are yet absent in the static limit
of the system. Therefore, the periodic driving allows us create multiple
gap-closing/reopening transitions in the quasienergy spectrum of the Floquet
system.

In Figs.~\ref{fig:C1GF}(b) and \ref{fig:C1GF}(c), we report the
Floquet spectrum of the system at different hopping amplitudes $J$
under the PBCs along both $x,y$ directions
(denoted by PBCXY) and under the open (periodic) boundary condition
along $x$ (y) direction (denoted by OBCX, PBCY). The gap-closing
points at $E=0$ and $\pm\pi$ in the bulk spectrum are found to be
the same as those observed in Fig.~\ref{fig:C1GF}(a). Meanwhile,
some curves representing the dispersion of edge bands are observed
in Fig.~\ref{fig:C1GF}(c) due to the open boundary condition taken
along the $x$-direction. Note that the quasienergies of edge states
in Fig.~\ref{fig:C1GF}(c) are not equal to zero and $\pm\pi$, even
though they look very close to them in some parameter regions. In
Fig.~\ref{fig:C1GF}(d), we present the quasienergies under the periodic
(open) boundary condition along $x$ ($y$) direction of the lattice
(denoted by PBCX, OBCY). Interestingly, despite the bulk gap-closing
points already seen in Fig.~\ref{fig:C1GF}(a), we observe another
gap-closing point at $E=0$ around $J=0$. This point goes beyond
the prediction of the bulk phase boundary according to Eq.~(\ref{eq:C1PB}).
As we will soon notice, this point corresponds to a closing of the
quasienergy gap between different edge bands under the given boundary
condition. It will thus be related to a type-II topological phase
transition \cite{HOTPT7} in our Floquet SOTSC system.

\begin{figure}
	\begin{centering}
		\includegraphics[width=10.5 cm]{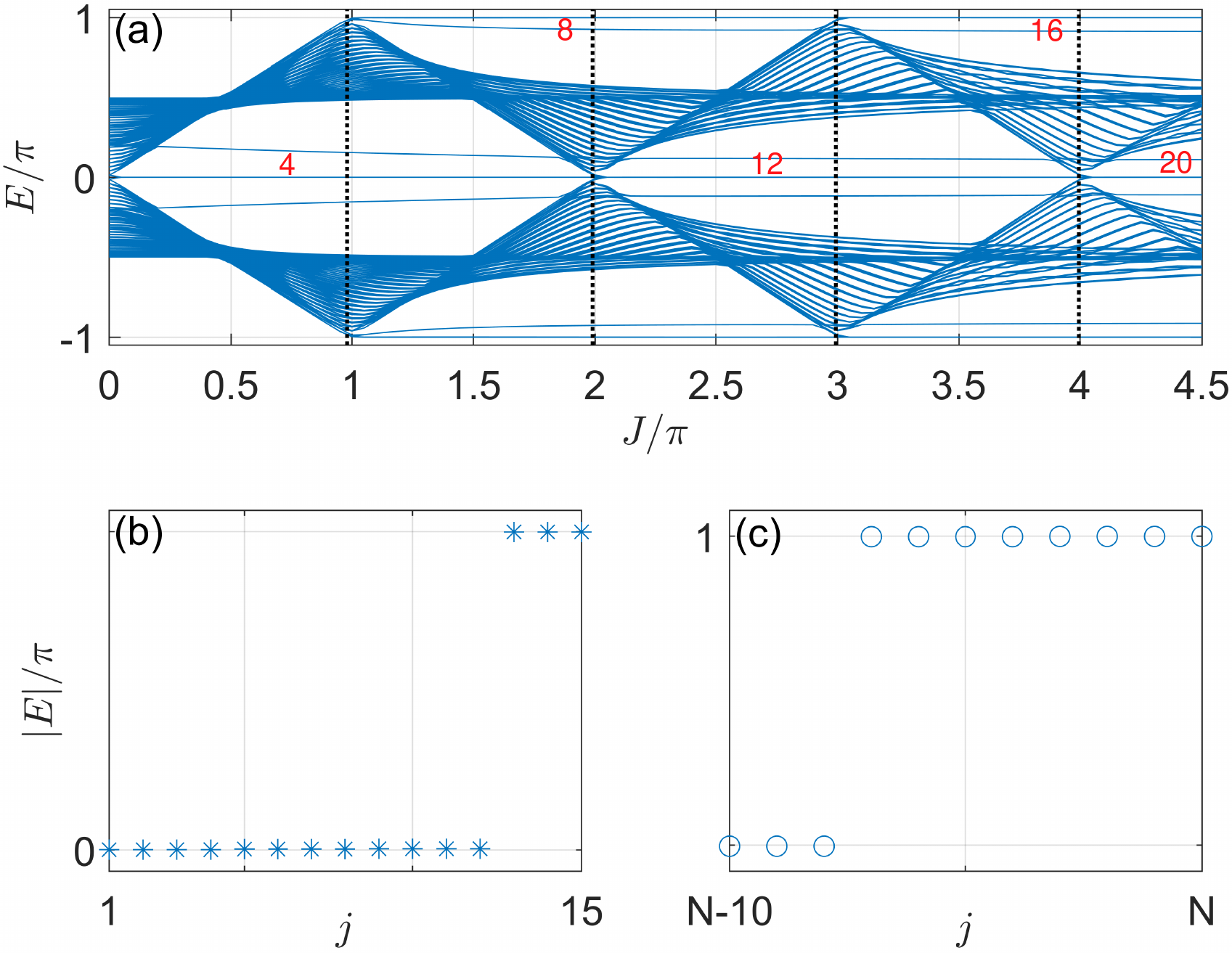}
		\par\end{centering}
	\caption{Floquet spectrum versus $J$ under OBCs in the Case 1. (a) Quasienergies
		at different $J$ under OBCs along both $x$ and $y$ directions (OBCXY).
		The crossing points between the vertical dotted lines and the horizontal
		axis show the bulk gap closing points predicted by Eq.~(\ref{eq:C1PB}).
		The numbers in red color denote the numbers of Floquet corner modes
		at zero and $\pi$ quasienergies. (b) and (c) show the absolute values
		of quasienergies of the first fifteen and last eleven Floquet eigenstates
		indexed by $j$ at $J=2.5\pi$. Other system parameters are $(\Delta,\Delta_{2})=(0.5\pi,0.2\pi)$.
		The number of cells along $x$ and $y$ directions are $N_{x}=N_{y}=60$.
		The total number of Floquet eigenstates is $N=14400$. \label{fig:C1EOBC}}
\end{figure}

\begin{figure}
	\begin{centering}
		\includegraphics[width=10.5 cm]{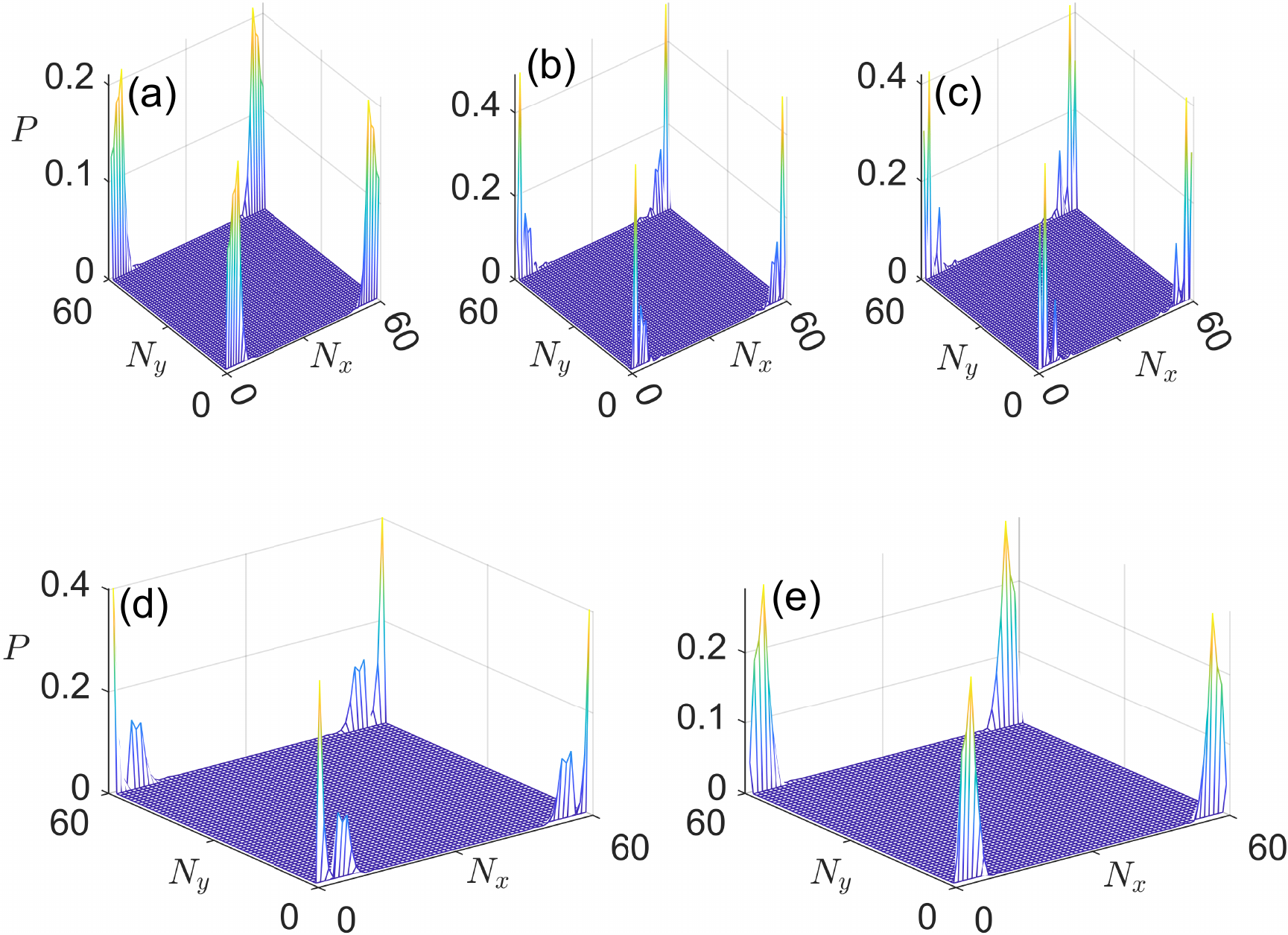}
		\par\end{centering}
	\caption{Probability distributions of Floquet corner modes in the Case 1 with
		zero and $\pi$ quasienergies in panels (a)--(c) and panels (d)--(e),
		respectively. Other system parameters are chosen to be $(J,\Delta,\Delta_{2})=(2.5\pi,0.5\pi,0.2\pi)$.
		The number of cells along $x$ and $y$ directions are $N_{x}=N_{y}=60$.
		\label{fig:C1POBC} }
\end{figure}

To deepen our understanding of the gap-closing transitions observed
in Case 1, we show the Floquet spectrum of the system under the OBCs
along both $x$ and $y$ directions (denoted by OBCXY) in Fig.~\ref{fig:C1EOBC}.
We find that more and more eigenmodes with $E=0$ or $\pm\pi$ emerge
following each gap-closing transition of the bulk as predicted by
Eq. (\ref{eq:C1PB}). In addition, the numbers of these eigenmodes
are integer multiples of four, which strongly suggest that they are
Floquet Majorana zero/$\pi$ modes localized around the four corners
of the lattice. In Fig. \ref{fig:C1POBC}(a)--(e), we report the
probability distributions of these zero/$\pi$ modes in the lattice
for a typical case {[}$(J,\Delta,\Delta_{2})=(2.5\pi,0.5\pi,0.2\pi)${]},
with their quasienergies displayed in Figs. \ref{fig:C1EOBC}(b) and
\ref{fig:C1EOBC}(c). The results confirm that they are indeed Majorana
corner modes in the Floquet system. Compared with the Floquet SOTSC
phases found in Ref. \cite{FHOTSC2}, we could now obtain many quartets
of Majorana zero and $\pi$ corner modes, which may give more room
for the realization of topological qubits and the operation of Floquet
quantum computing protocols as suggested previously \cite{FHOTSC2}.
Besides, with the increase of $J$, we could in principle obtain unbounded
numbers of Floquet zero/$\pi$ Majorana corner modes in the thermodynamic
limit. This demonstrates again one key advantage of Floquet engineering,
i.e., to generate many topological nontrivial states and topological
phase transitions in a controlled manner. 

\begin{figure}
	\begin{centering}
		\includegraphics[width=10.5 cm]{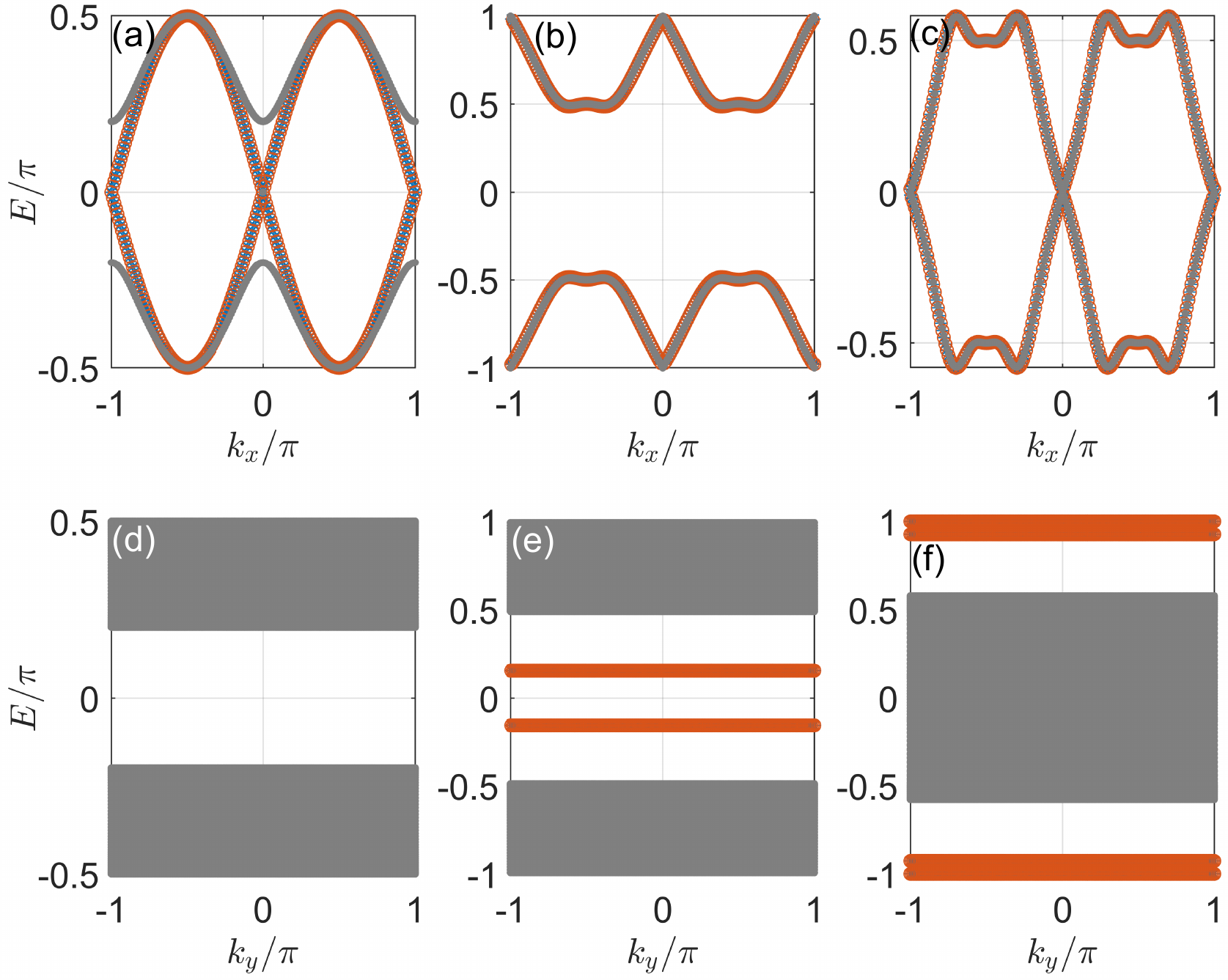}
		\par\end{centering}
	\caption{Floquet spectrum versus $k_{x}$ ($k_{y}$) in the Case 1 under PBCX,
		OBCY (OBCX, PBCY) in panels (a)--(c) {[}(d)--(f){]}. The gray dots,
		red circles and blue stars highlight the bulk states, states localized
		around the left edge and the right edge of the lattice. The value
		of hopping amplitude is set to $J=0$ for panels (a), (d), to the
		first bulk gap-closing point at $E=\pm\pi$ for panels (b), (e), and
		to the first bulk gap-closing point at $E=0$ for panels (c), (f).
		Other system parameters are set as $(\Delta,\Delta_{2})=(0.5\pi,0.2\pi)$.
		\label{fig:C1EMBC}}
\end{figure}

Finally, we are left to understand the transition at $J=0$ in 
Fig.~\ref{fig:C1EOBC}(a), following which four Floquet Majorana corner
modes appear at $E=0$. However, the bulk Floquet spectral gap remains
open throughout this transition, as observed clearly in Fig.~\ref{fig:C1GF}(b).
By diagonalizing the $\hat{U}(k_{x})$ in Eq. (\ref{eq:Ukx}), we
obtain the Floquet spectrum of the system versus $k_{x}$ under the
PBCX, OBCY, as reported in Figs. \ref{fig:C1EMBC}(a)--(c) for $J=0$,
$J=\sqrt{\pi^{2}-\Delta_{2}^{2}}$ and $J=\sqrt{(2\pi)^{2}-\Delta_{2}^{2}}$,
respectively. We observe that while the bulk gaps close at $E=\pm\pi$
and $E=0$ in the latter two cases, only edge states develop crossings
at $E=0$ when $J=0$. For comparison, we obtain the Floquet spectrum
versus the quasimomentum $k_{y}$ under the OBCX, PBCY by diagonalizing
the $\hat{U}(k_{y})$ in Eq. (\ref{eq:Uky}), as reported in Figs.
\ref{fig:C1EMBC}(d)--(f). We again find the closings of bulk spectrum
gaps at $E=\pm\pi$ and $E=0$ for $J=\sqrt{\pi^{2}-\Delta_{2}^{2}}$
and $J=\sqrt{(2\pi)^{2}-\Delta_{2}^{2}}$, respectively. These are
the bulk transition points predicted by Eq. (\ref{eq:C1PB}). Yet,
the spectrum is found to be well gapped at $E=0$ for $J=0$ in Figs.
\ref{fig:C1EMBC}(d). Putting together, we conclude that under the condition
in Eq. (\ref{eq:C1PB}), Floquet zero/$\pi$ corner modes emerge following
type-I topological phase transitions (with bulk-band touchings) in
the Case 1. Meanwhile, the Floquet zero corner modes could also appear
following a type-II higher-order topological phase transition (with
edge-band touchings) at $J=0$. There are thus two types of topological
phase transitions in our Floquet SOTSC system, with each of them being
able to generate more Floquet Majorana corner modes. This is another
key difference between our system and that explored before \cite{FHOTSC2}.

The model presented in the Case 1 thus forms a ``minimal'' model
of Floquet second-order topological superconductors with rich and
different types of topological phase transitions, along with unbounded
numbers of Floquet Majorana zero/$\pi$ corner modes. All of them
are induced by a simple time-periodic driving protocol applied to
the superconducting pairing amplitude. In the following subsections,
we explore more general situations and check how the introducing of
extra onsite, pairing and hopping terms could affect the corner modes
and the phase transitions found in our system.

\subsection{Case 2: $J'=\Delta_1=0$}\label{subsec:C2}

In a slightly more general situation, we allow the system to possess
a finite chemical potential $\mu$. From Eqs. (\ref{eq:Uk})--(\ref{eq:H2k}),
the resulting Floquet operator in momentum space now takes the form

\begin{equation}
	U({\bf k})=e^{-i[(\mu+J\cos k_{x})\sigma_{0}\otimes\sigma_{z}+\Delta_{2}(\sin k_{y}\sigma_{x}-\cos k_{y}\sigma_{y})\otimes\sigma_{x}]}
	e^{-i\Delta\sin k_{x}\sigma_{0}\otimes\sigma_{y}}.\label{eq:C2U}
\end{equation}
Similar to the Case 1, the tensor product matrices $\sigma_{0}\otimes\sigma_{z}$,
$\sigma_{x}\otimes\sigma_{x}$, $\sigma_{y}\otimes\sigma_{x}$ and
$\sigma_{0}\otimes\sigma_{y}$ are anti-commute with one another.
Therefore, we can obtain the bulk quasienergy dispersion relation as

\begin{equation}
	E_{\pm}({\bf k})=\pm\arccos\left[\cos(\Delta\sin k_{x})\cos\left(\sqrt{(\mu+J\cos k_{x})^{2}+\Delta_{2}^{2}}\right)\right].\label{eq:C2E}
\end{equation}
There are again two pairs of quasienergy bands with each of them been
twofold degenerate and flat along $k_y$. Setting $E_{\pm}({\bf k})=0$ or $\pm\pi$, we
find the gapless condition of Floquet spectrum to be

\begin{equation}
	\cos(\Delta\sin k_{x})\cos\left(\sqrt{(\mu+J\cos k_{x})^{2}+\Delta_{2}^{2}}\right)=\pm1,\label{eq:C2PB0}
\end{equation}
yielding the phase boundary equation in the parameter space $(\Delta,\mu,J,\Delta_{2})$ as

\begin{equation}
	\frac{p^{2}\pi^{2}}{\Delta^{2}}+\frac{\left(\sqrt{q^{2}\pi^{2}-\Delta_{2}^{2}}\pm\mu\right)^{2}}{J^{2}}=1,\qquad p,q\in\mathbb{Z}.\label{eq:C2PB}
\end{equation}
In contrast to the Eq.~(\ref{eq:C1PB}), we see that the general effect
of a nonvanishing $\mu$ is to split each transition point of the
Case 1 into two distinct points separated by a distance $\sim2\mu$
on the phase diagram. We would thus expect richer gap closing/reopening
transitions in the parameter space of Case $2$ compared with the
Case 1. Meanwhile, the gap functions in Case 2 share the same forms
with Eq.~(\ref{eq:F0P}) of the Case 1, as the
systems possess two pairs of twofold-degenerate Floquet bands in both
cases.

\begin{figure}
	\begin{centering}
		\includegraphics[width=10.5 cm]{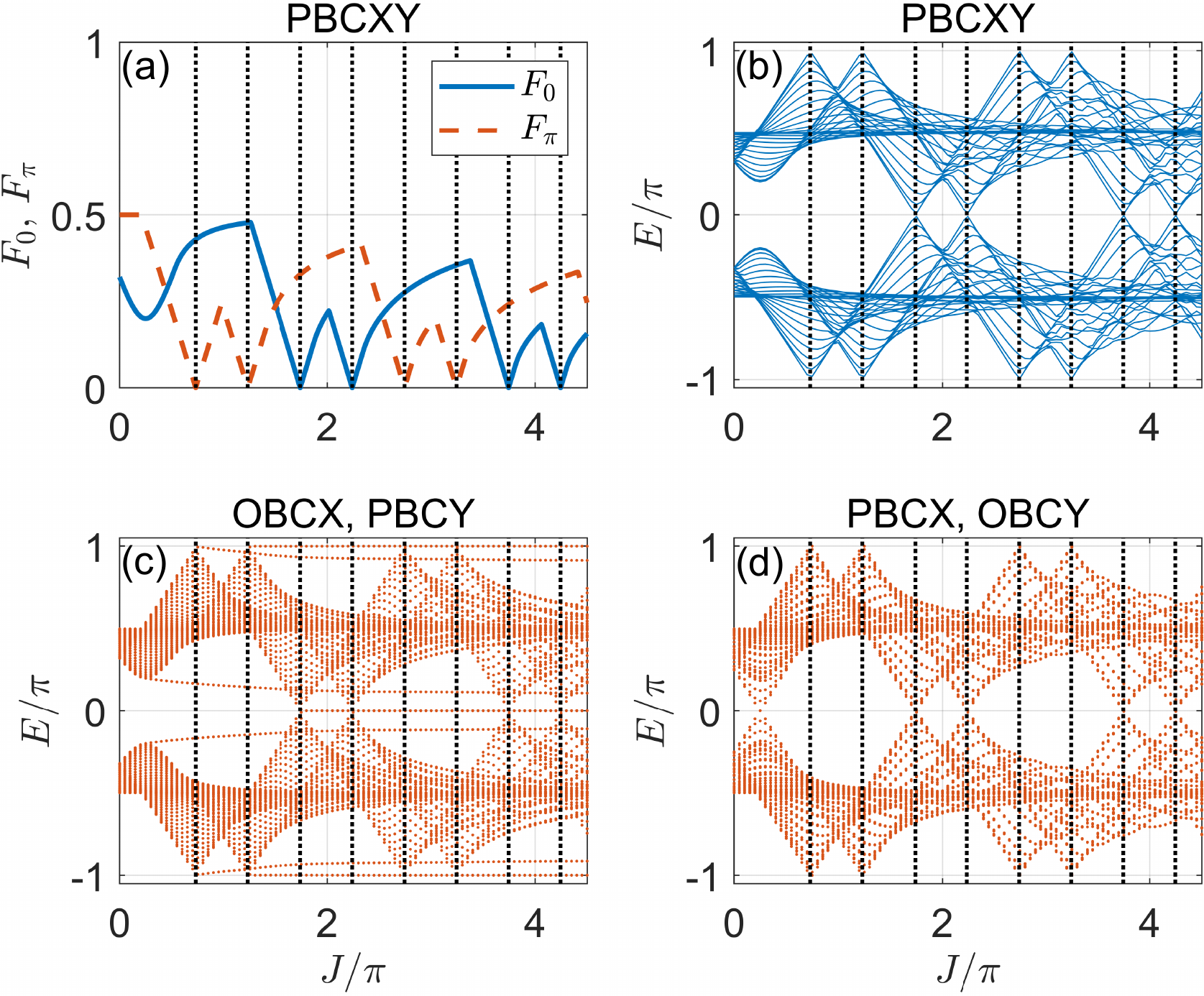}
		\par\end{centering}
	\caption{Gap functions and Floquet spectrum versus $J$ under different boundary
		conditions in the Case 2. (a) Gap functions under PBCXY. (b) Floquet
		spectrum under PBCXY. (c) Floquet spectrum under OBCX, PBCY. (d) Floquet
		spectrum under PBCX, OBCY. The crossing points between the vertical
		dotted lines and the horizontal axis show the bulk gap closing points
		predicted by Eq.~(\ref{eq:C2PB}). Other system parameters are set
		as $(\mu,\Delta,\Delta_{2})=(0.25\pi,0.5\pi,0.2\pi)$. The number
		of cells along $x$ and $y$ directions are $N_{x}=N_{y}=60$. \label{fig:C2GF}}
\end{figure}

In Fig.~\ref{fig:C2GF}(a), we present the gap functions $F_{0}$
and $F_{\pi}$ versus $J$ for a typical set of system parameters
in Case 2. The first interesting observation is that each gap-closing
transition point located originally at $J=\sqrt{q^{2}\pi^{2}-\Delta_{2}^{2}}$
(for $p=0$, $q\in\mathbb{Z}$) in Fig.~\ref{fig:C1GF}(a) now splits
into two points residing at $J=\sqrt{q^{2}\pi^{2}-\Delta_{2}^{2}}\pm\mu$
in Fig.~\ref{fig:C2GF}(a), as also predicted by the phase boundary
Eq.~(\ref{eq:C2PB}). Therefore, the presence of a finite chemical
potential $\mu$ endues the system with richer transition patterns
in its spectrum. This is further confirmed by the quasienergies with
respect to $J$ under three different boundary conditions, as reported
in Figs.~\ref{fig:C2GF}(b)--(d). The bulk gap-closing points in
each figure are found to be consistent with the prediction of Eq.
(\ref{eq:C2PB}). Notably, we observe in Fig.~\ref{fig:C2GF}(d) a
gapless point in the Floquet spectrum at $J=\mu$, which is not captured
by Eq.~(\ref{eq:C2PB}). We will see that this point is again related
to a touching of the edge bands instead of the bulk spectrum, similar
to what we have encountered at $J=0$ in the Case 1. 

\begin{figure}
	\begin{centering}
		\includegraphics[width=10.5 cm]{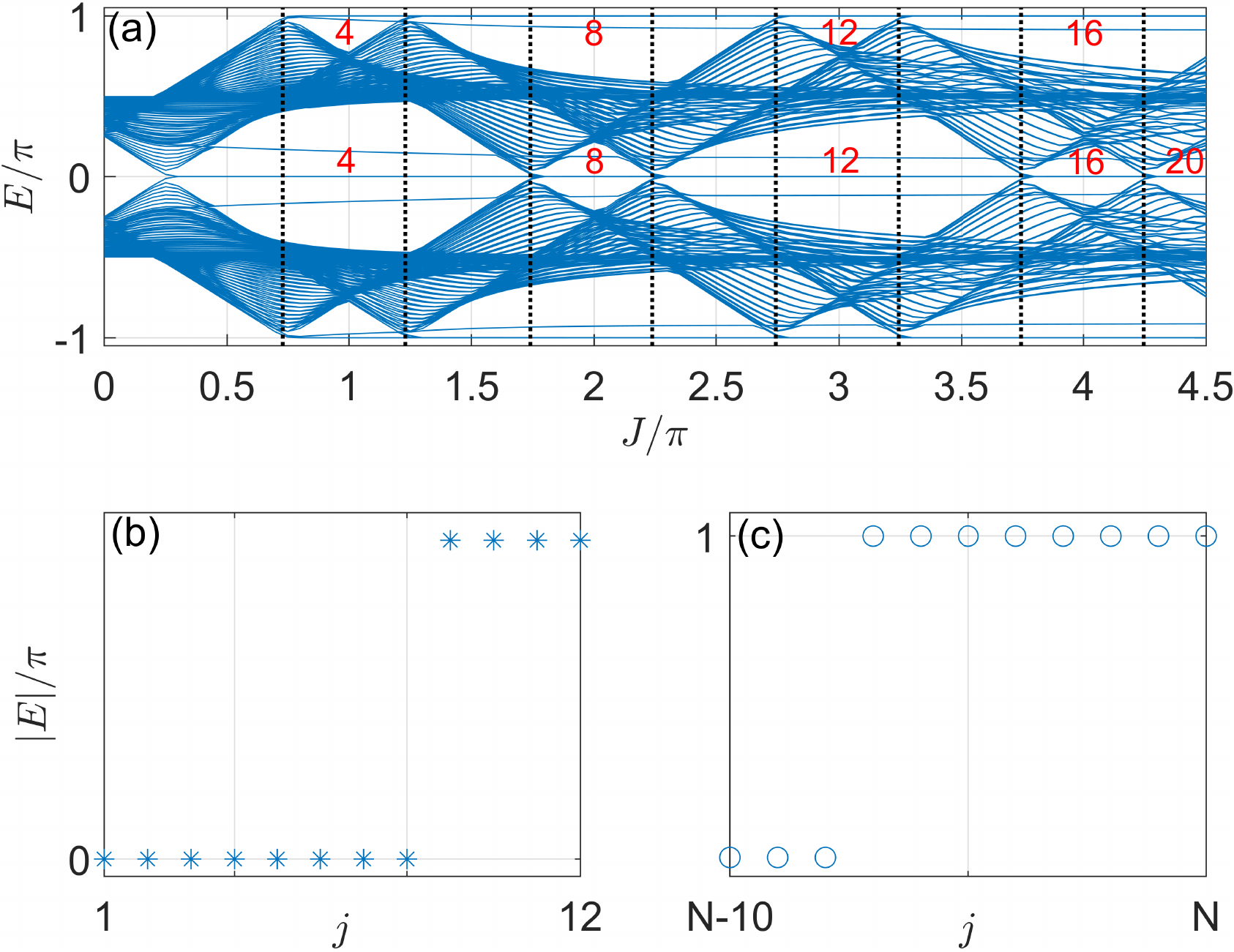}
		\par\end{centering}
	\caption{Floquet spectrum versus $J$ under OBCs in the Case 2. (a) Quasienergies
		at different $J$ under OBCXY. The crossing points between the vertical
		dotted lines and the horizontal axis show the bulk gap-closing points
		predicted by the Eq.~(\ref{eq:C2PB}). The numbers in red color denote
		the numbers of Floquet corner modes at zero and $\pi$ quasienergies.
		(b) and (c) show the absolute values of quasienergies of the first
		twelve and last eleven Floquet eigenstates indexed by $j$ at $J=2\pi$.
		Other system parameters are $(\mu,\Delta,\Delta_{2})=(0.25\pi,0.5\pi,0.2\pi)$.
		The number of cells along $x$ and $y$ directions are $N_{x}=N_{y}=60$.
		The total number of Floquet eigenstates is $N=14400$. \label{fig:C2EOBC}}
\end{figure}

The spectrum of our system in the Case 2 under OBCs is shown in 
Fig.~\ref{fig:C2EOBC}. We observe that following each gap-closing transition,
there are indeed more eigenmodes emerge at the quasienergy zero or
$\pi$. Their numbers in each phase are denoted by the numbers in
red in Fig. \ref{fig:C2EOBC}(a). Similar to the Case 1, here the
numbers of zero and $\pi$ Floquet eigenmodes are both integer multiples
of four, implying that they are fourfold degenerate Majorana corner
states in Floquet SOTSC phases. However, across each bulk gap-closing
transition, the number of Floquet zero or $\pi$ eigenmodes now only
changes by $4$ instead of $8$. This is originated from the splitting
of each gapless point into two in the phase diagram with a nonvanishing
$\mu$. Following each of the new generated transitions, the change
of Floquet zero/$\pi$ eigenmode number is just half of the original
transition in the Case 1. We will further digest this point by investigating
the momentum space spectrum of the system later in this subsection.
Besides, we observe monotonic increases of the numbers of zero and
$\pi$ Floquet eigenmodes with the increase of $J$ following the
consecutive spectrum transitions. It implies that we could also obtain
a great deal of Floquet zero and $\pi$ corner modes at large $J$
in the thermodynamic limit when $\mu\neq0$. Meanwhile, four Floquet
zero corner modes emerge from a transition at $J=\mu$, which is not
captured by the bulk phase boundary Eq.~(\ref{eq:C2PB}). As mentioned
before, we will trace it back to the touching of edge state bands,
i.e., a type-II topological phase transition.

\begin{figure}
	\begin{centering}
		\includegraphics[width=10.5 cm]{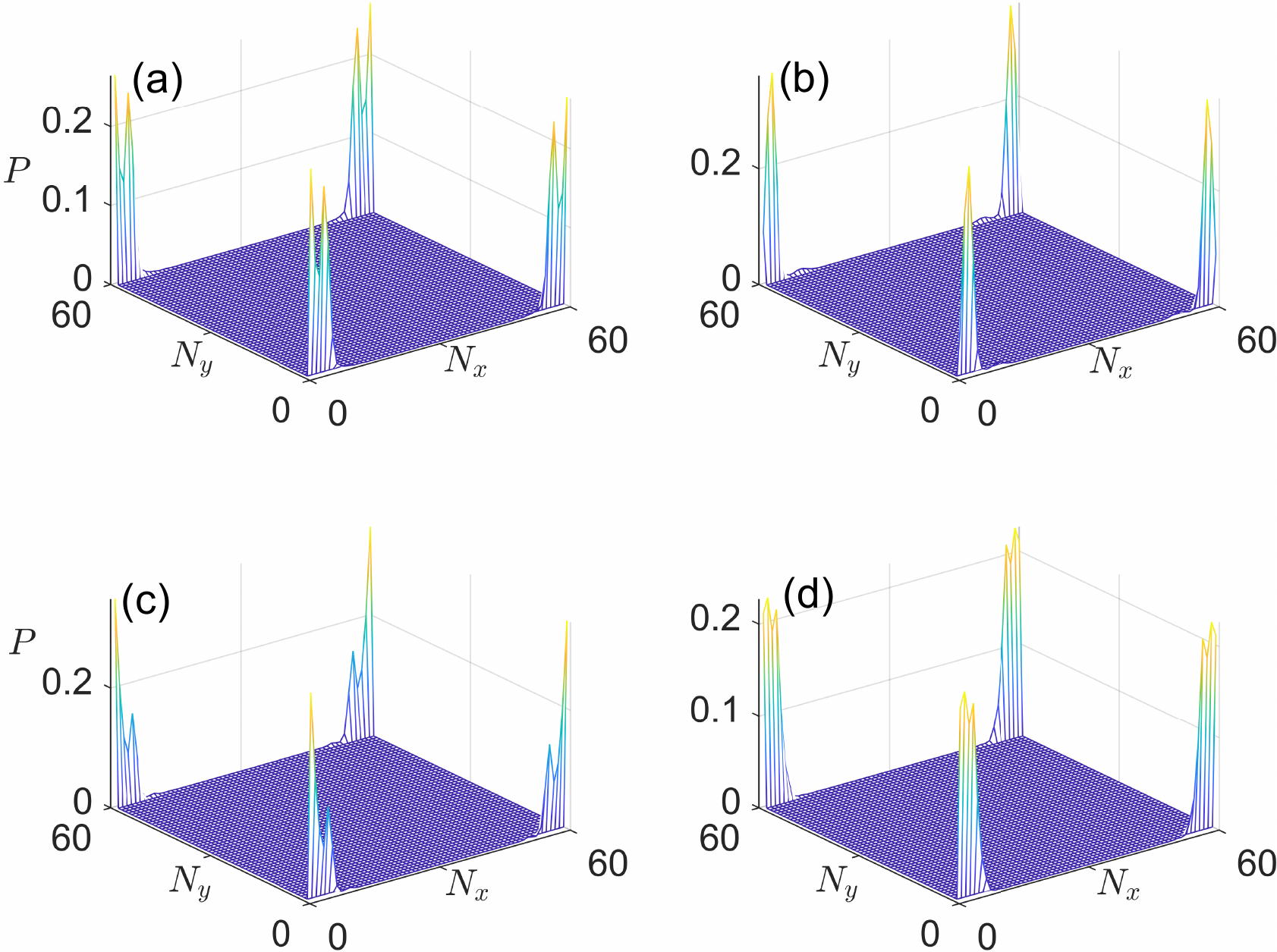}
		\par\end{centering}
	\caption{Probability distributions of Floquet corner modes in the Case 2 with
		quasienergies zero and $\pi$ in the panels (a)--(b) and panels (c)--(d),
		respectively. Other system parameters are set as $(J,\mu,\Delta,\Delta_{2})=(2\pi,0.25\pi,0.5\pi,0.2\pi)$.
		The number of cells along $x$ and $y$ directions of the lattice
		are $N_{x}=N_{y}=60$. \label{fig:C2POBC}}
\end{figure}

To confirm that the zero and $\pi$ Floquet eigenmodes observed in
the spectrum are indeed Majorana corner modes, we plot their probability
distributions in the lattice for a typical choice of system parameters
in Fig. \ref{fig:C2POBC}. The plotted profiles correspond to the
first eight and the last eight eigenstates shown in Figs. \ref{fig:C2EOBC}(b)
and \ref{fig:C2EOBC}(c). The numbers of zero and $\pi$ eigenmodes
are $(n_{0},n_{\pi})=(8,8)$ in this case according to Fig. \ref{fig:C2EOBC}(a),
and we see that they are indeed localized around the four corners
of the lattice. Therefore, we could obtain many quartets of
zero and $\pi$ Majorana corner modes in our Floquet SOTSC system
even with $\mu\neq0$. This goes beyond the situation encountered
before, where only four Majorana zero and $\pi$ corner modes are
found to coexist in a continuously driven system \cite{FHOTSC2}.
The many corner Majorana states discovered here may again give us
more freedoms to implement Floquet quantum computing schemes \cite{FHOTSC2}.

\begin{figure}
	\begin{centering}
		\includegraphics[width=10.5 cm]{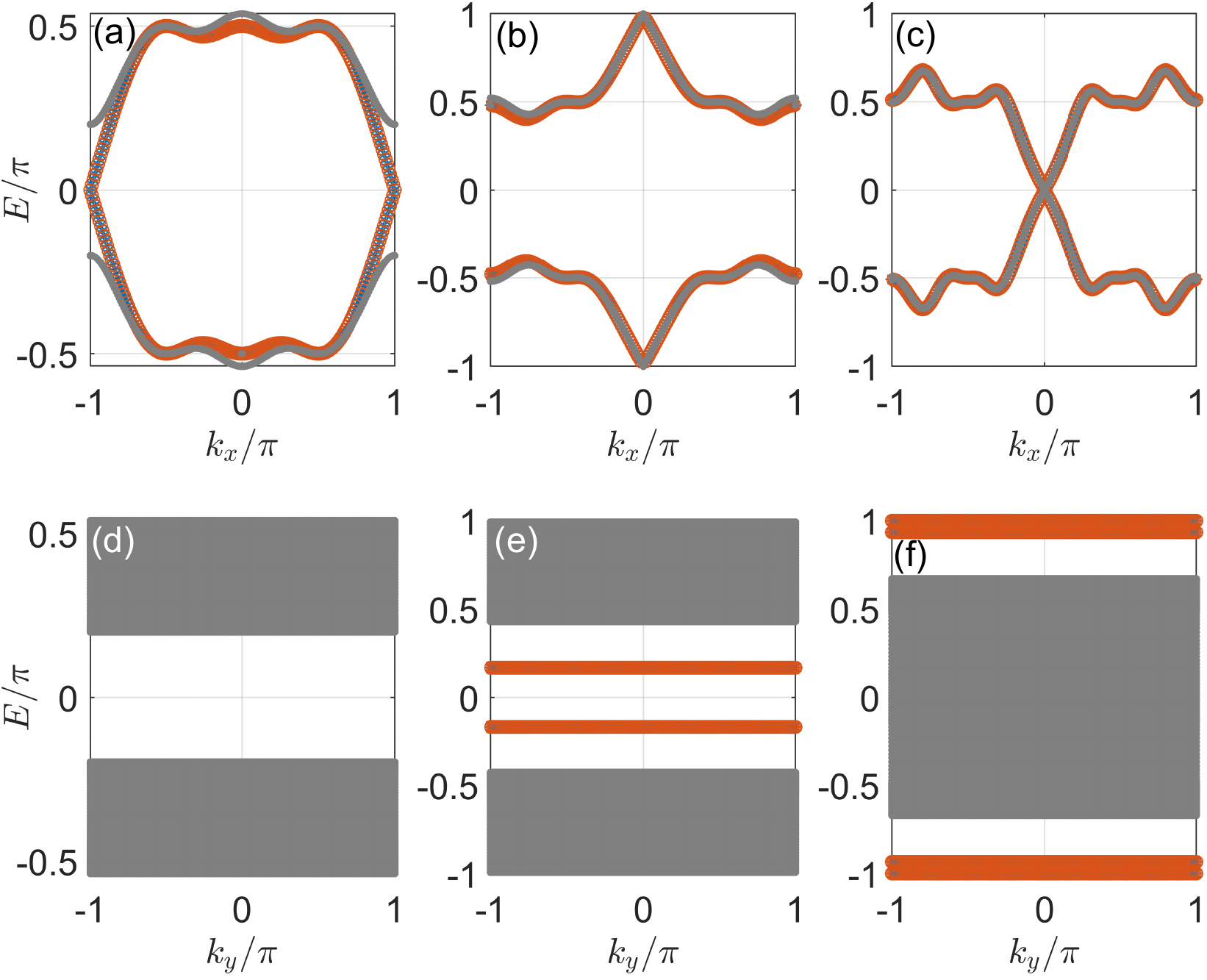}
		\par\end{centering}
	\caption{Floquet spectrum versus $k_{x}$ ($k_{y}$) in the Case 2 under PBCX,
		OBCY (OBCX, PBCY) in panels (a)--(c) {[}(d)--(f){]}. The gray dots,
		red circles and blue stars highlight the bulk states, states localized
		around the left edge and the right edge of the lattice. The value
		of hopping amplitude is set to $J=\mu$ for the panels (a), (d), to
		the first bulk gap-closing point at $E=\pm\pi$ for the panels (b),
		(e), and to the first bulk gap-closing point at $E=0$ for the panels
		(c), (f). Other system parameters are chosen to be $(\mu,\Delta,\Delta_{2})=(0.25\pi,0.5\pi,0.2\pi)$.
		\label{fig:C2EMBC}}
\end{figure}

To understand the difference between the anomalous transition observed
at $J=\mu$ and the other bulk gap-closing transitions, we present
the spectra of the system under the PBCX, OBCY versus $k_{x}$ and
under the OBCX, PBCY versus $k_{y}$ at three critical points $J=\mu$,
$J=\sqrt{\pi^{2}-\Delta_{2}^{2}}-\mu$ and $J=\sqrt{(2\pi)^{2}-\Delta_{2}^{2}}-\mu$
in Fig. \ref{fig:C2EMBC}. These spectra are obtained by diagonalizing
$\hat{U}(k_{x})$ and $\hat{U}(k_{y})$ in Eqs. (\ref{eq:Ukx}) and
(\ref{eq:Uky}), respectively. We find that for the transitions at
$J=\sqrt{\pi^{2}-\Delta_{2}^{2}}-\mu$ and $J=\sqrt{(2\pi)^{2}-\Delta_{2}^{2}}-\mu$,
the bulk Floquet bands of the system indeed touch at $E=\pm\pi$ and
$E=0$ in Figs. \ref{fig:C2EMBC}(b), \ref{fig:C2EMBC}(e) and \ref{fig:C2EMBC}(c),
\ref{fig:C2EMBC}(f). This is consistent with the prediction of the
bulk phase boundary in Eq. (\ref{eq:C2PB}). Besides, a comparison
between the Figs. \ref{fig:C2EMBC}(b)--(c) and Figs. \ref{fig:C1EMBC}(b)--(c)
also helps us to understand why the change of zero/$\pi$ corner mode
numbers is four instead of eight across each topological transition
when $\mu\neq0$. That is, in the Case 1 the Floquet bands touch at
two distinct points along $k_{x}$ {[}$k_{x}=0,\pi$ in Figs. \ref{fig:C1EMBC}(b)--(c){]}
at the bulk phase transition point. Meanwhile, in the Case 2 the Floquet
bands only touch at a single point along $k_{x}$ {[}$k_{x}=0$ in
Figs. \ref{fig:C2EMBC}(b)--(c){]} at each bulk phase transition
point, yielding a smaller change in the number of Majorana corner
modes across the transition.

Finally, we realize that at $J=\mu$, the edge state bands develop
a touching at $k_{x}=0$ under the PBCX, OBCY in Fig.~\ref{fig:C2EMBC}(a),
while no edge states and band touchings are observed under the OBCX,
PBCY in Fig.~\ref{fig:C2EMBC}(d). This observation confirms that
the four Floquet Majorana zero corner modes indeed emerge out of a
type-II topological phase transition mediated by an edge-band touching
at $J=\mu$ in Fig.~\ref{fig:C2POBC}(a). Therefore, we also encounter
two types of topological phase transitions in the Case 2, after which
more Floquet Majorana zero/$\pi$ corner modes appear with the increase
of the hopping amplitude $J$ in our system. This tendency will continue
in more general situations as will be discussed in the following subsection.

\subsection{Case 3: $J'=0$}\label{subsec:C3}

We now consider the case with only $J'=0$. In this case, all the
pairing terms $(\Delta,\Delta_{1},\Delta_{2})$ are switched on. Following
Eqs. (\ref{eq:Uk})--(\ref{eq:H2k}), the Floquet operator in momentum
space takes the form

\begin{equation}
	U({\bf k})= e^{-i\{(\mu+J\cos k_{x})\sigma_{0}\otimes\sigma_{z}+[\Delta_{2}\sin k_{y}\sigma_{x}+(\Delta_{1}-\Delta_{2}\cos k_{y})\sigma_{y}]\otimes\sigma_{x}\}} e^{-i\Delta\sin k_{x}\sigma_{0}\otimes\sigma_{y}}.\label{eq:C3U}
\end{equation}
Using again the anticommuting nature of tensor product matrices $\sigma_{0}\otimes\sigma_{z}$,
$\sigma_{x}\otimes\sigma_{x}$, $\sigma_{y}\otimes\sigma_{x}$, $\sigma_{0}\otimes\sigma_{y}$
and the Taylor expansion, we find the quasienergy dispersions of $U({\bf k})$ to be

\begin{equation}
	E_{\pm}({\bf k})=\pm\arccos\left[\cos(\Delta\sin k_{x})\cos\left(\sqrt{(\mu+J\cos k_{x})^{2}+\Delta_{1}^{2}+\Delta_{2}^{2}-2\Delta_{1}\Delta_{2}\cos k_{y}}\right)\right].\label{eq:C3E}
\end{equation}
As before, we obtain two pairs of twofold-degenerate Floquet bands.
However, the quasienergies $E_{\pm}({\bf k})$ now depend on both
the quasimomenta $k_{x}$ and $k_{y}$ due to the nonvanishing $\Delta_{1}$.
Setting $E_{\pm}({\bf k})$ to zero or $\pm\pi$, we obtain the gapless
condition of the spectrum as

\begin{equation}
	\cos(\Delta\sin k_{x})\cos\left(\sqrt{(\mu+J\cos k_{x})^{2}+\Delta_{1}^{2}+\Delta_{2}^{2}-2\Delta_{1}\Delta_{2}\cos k_{y}}\right)=\pm1.\label{eq:C3PB0}
\end{equation}
Without loss of generality, we now assume all system parameters $(\mu,J,\Delta,\Delta_{1},\Delta_{2})$
in the Case $3$ to be positive. One can then deduce from Eq. (\ref{eq:C3PB0})
that the gap between the two sets of Floquet bands vanishes at $E=0$ or $\pi$ when the inequality

\begin{equation}
	q^{2}\pi^{2}-(\Delta_{1}+\Delta_{2})^{2}\leq\left(\mu\pm J\sqrt{1-\frac{p^{2}\pi^{2}}{\Delta^{2}}}\right)^{2}\leq q^{2}\pi^{2}-(\Delta_{1}-\Delta_{2})^{2}\label{eq:C3PB}
\end{equation}
is satisfied for $p,q\in\mathbb{Z}$. This relation suggests a rather
different situation compared with the Cases 1 and 2, i.e., the Floquet
spectrum could now become gapless at the quasienergy zero or $\pi$
in finite domains of the parameter space. Therefore, our model in
the Case $3$ could possess gapless phases with semimetal-like Floquet
band structures. Two different Floquet SOTSC phases may then be separated
by a finite gapless region instead of a gap-closing point in the spectrum.
In the meantime, we can still define the gap functions as the 
Eq.~(\ref{eq:F0P}) in Case 1.

\begin{figure}
	\begin{centering}
		\includegraphics[width=10.5 cm]{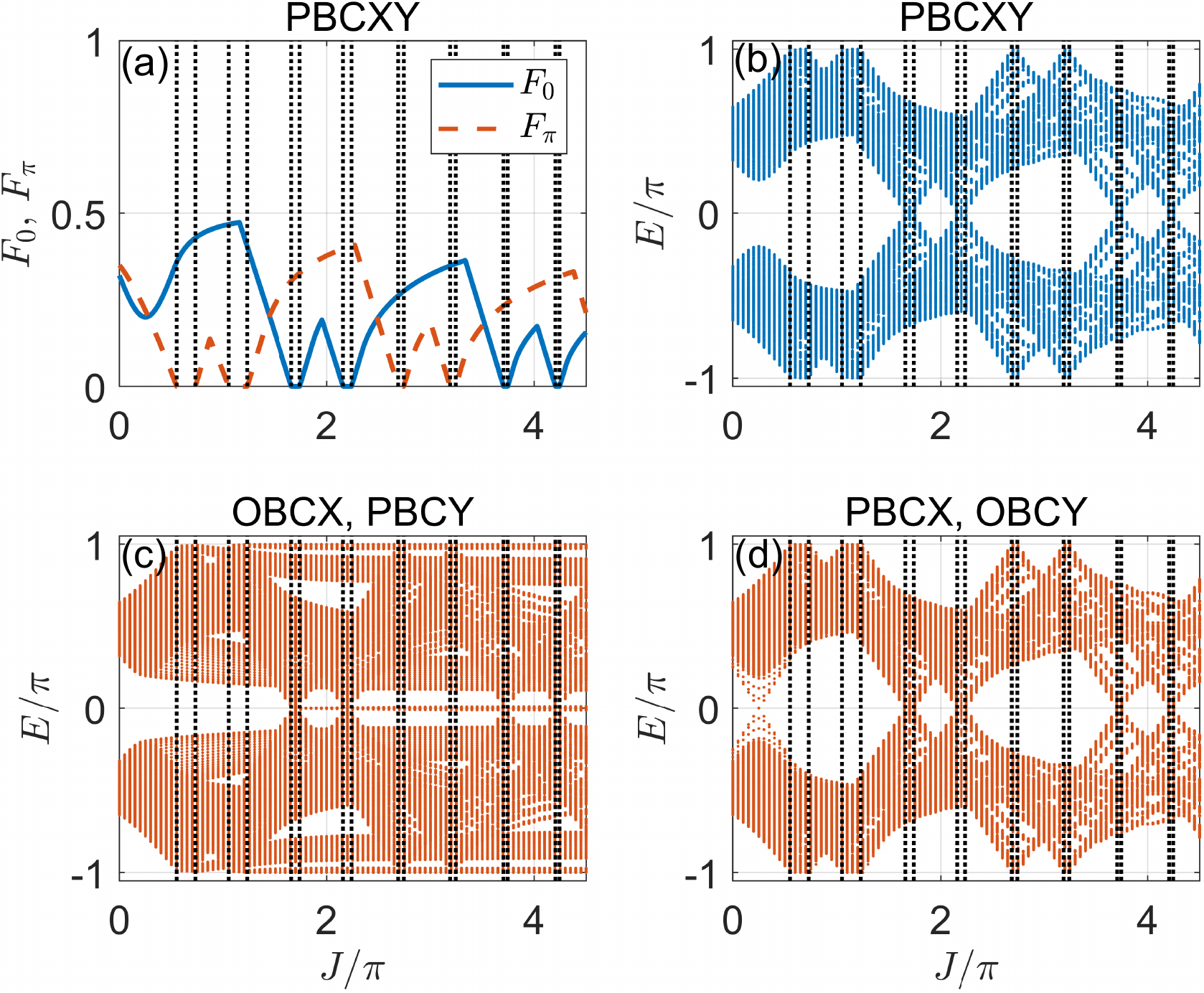}
		\par\end{centering}
	\caption{Gap functions and Floquet spectrum versus $J$ under different boundary
		conditions in the Case 3. (a) Gap functions under PBCXY. (b) Floquet
		spectrum under PBCXY. (c) Floquet spectrum under OBCX, PBCY. (d) Floquet
		spectrum under PBCX, OBCY. The crossing points between the vertical
		dotted lines and the horizontal axis show the bulk gap-closing points
		predicted by the Eq.~(\ref{eq:C3PB}). Other system parameters are
		set as $(\mu,\Delta,\Delta_{1},\Delta_{2})=(0.25\pi,0.5\pi,0.2\pi,0.4\pi)$.
		The number of cells along $x$ and $y$ directions are $N_{x}=N_{y}=60$.
		\label{fig:C3GF}}
\end{figure}

In Fig.~\ref{fig:C3GF}(a), we plot the gap functions $F_{0}$ and
$F_{\pi}$ versus $J$ for a typical set of system parameters in the
Case 3. Interestingly, we observe that with a finite intracell pairing
$\Delta_{1}$, the gap-closing points in the Case 2 now broaden into
regions of finite widths along the $J$-axis. The boundaries of these
gapless bulk regions are highlighted by the vertical dotted lines
in Fig.~\ref{fig:C3GF}, which are further determined theoretically
by Eq.~(\ref{eq:C3PB}). Therefore, in the presence of a finite $\Delta_{1}$,
the transitions between different Floquet SOTSC phases can be mediated
by other phases with gapless Floquet spectra instead of isolated critical
points in the parameter space. The existence of these gapless phases
is further confirmed by the Floquet spectrum presented in Figs.~\ref{fig:C3GF}(b)--(d)
under different boundary conditions. We could thus obtain richer patterns
of Floquet phases and spectral transitions with a nonvanishing $\Delta_{1}$.
This is distinct from the situation considered previously, where the
gapless phases are not observed in a harmonically driven setting \cite{FHOTSC2}.
Despite the many transitions occurred with the increase of $J$, we
observe another isolated transition at $J=\mu$ that is not described
by Eq.~(\ref{eq:C3PB}). As discussed in previous subsections, this
transition is also expected to be originated from the touching of
edge state bands instead of bulk bands.

\begin{figure}
	\begin{centering}
		\includegraphics[width=10.5 cm]{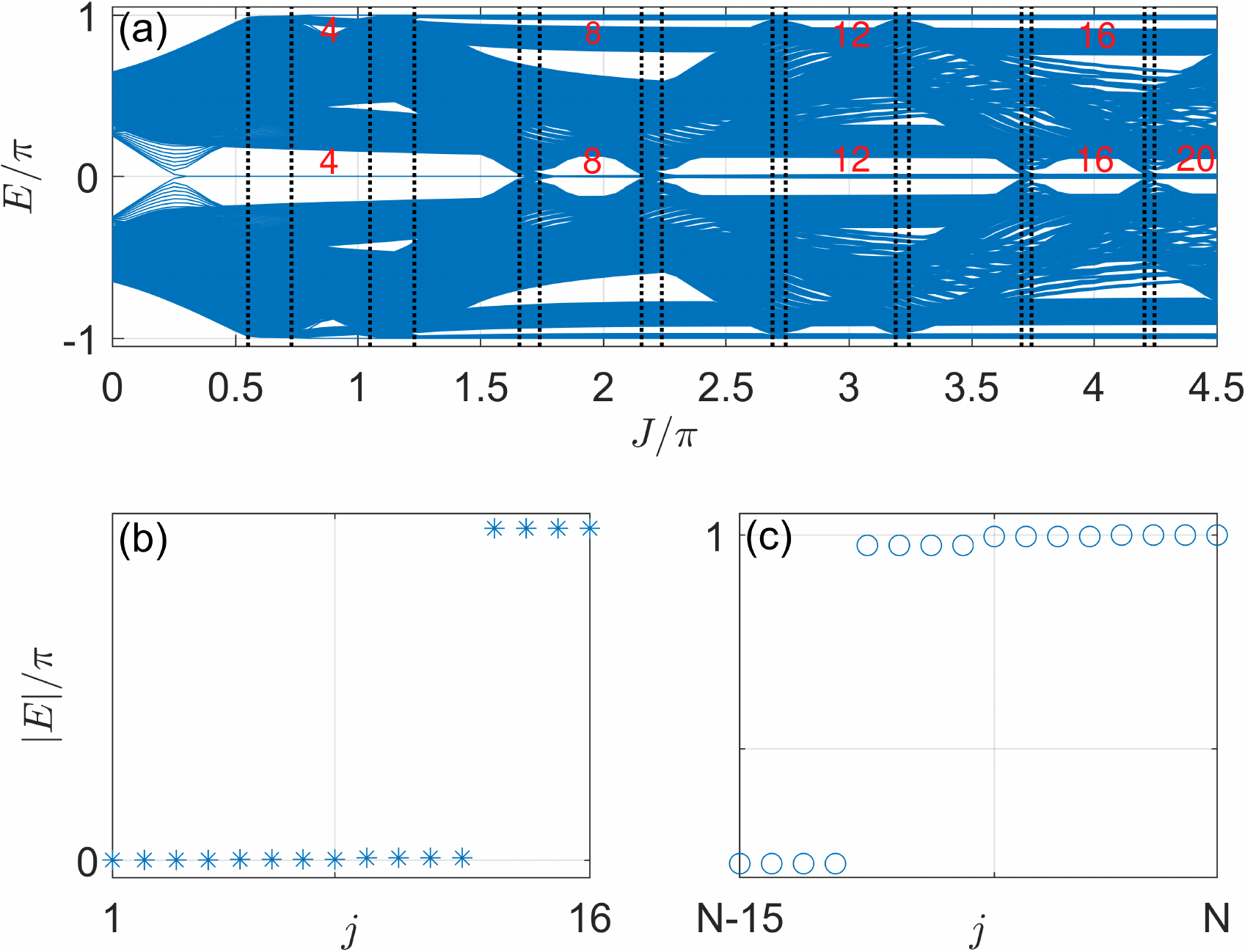}
		\par\end{centering}
	\caption{Floquet spectrum versus $J$ under OBCs in the Case 3. (a) Quasienergies
		at different $J$ under OBCXY. The crossing points between the vertical
		dotted lines and the horizontal axis show the bulk gap-closing points
		predicted by the Eq.~(\ref{eq:C3PB}). The numbers in red color denote
		the numbers of Floquet corner modes at zero and $\pi$ quasienergies.
		(b) and (c) show the absolute values of quasienergies of the first
		sixteen and last sixteen Floquet eigenstates indexed by $j$ at $J=3\pi$.
		Other system parameters are $(\mu,\Delta,\Delta_{1},\Delta_{2})=(0.25\pi,0.5\pi,0.2\pi,0.4\pi)$.
		The number of cells along $x$ and $y$ directions are $N_{x}=N_{y}=60$.
		The total number of Floquet eigenstates is $N=14400$. \label{fig:C3EOBC}}
\end{figure}

In Fig.~\ref{fig:C3EOBC}, we report the Floquet spectrum of the system
under OBCs along both directions. Following either the transition
at $J=\mu$ or the transition mediated by a gapless phase at $E=0$
or $E=\pm\pi$, we obtain more eigenmodes at the quasienergy zero
or $\pi$ with the increase of $J$, whose numbers are highlighted
in red in Fig.~\ref{fig:C3EOBC}(a). The $4\mathbb{Z}$-quantization
of these eigenmode numbers again suggest that they are localized states
around the four corners of the lattice. Similar to the previous two
cases, we could foresee unboundedly many quartets of eigenmodes at
$E=0$ and $\pm\pi$ with $J\rightarrow\infty$ in the thermodynamic
limit. This demonstrates a controllable generation of these topological
modes in our Floquet SOTSC model under a simple driving protocol.
The main impact of a finite pairing amplitude $\Delta_{1}\neq0$ is
thus to generate gapless regions between different Floquet SOTSC phases,
while preserving the zero and $\pi$ Floquet eigenmodes in the gapped
phases of the bulk. Besides, four Floquet zero modes emerge through
the transition at $J=\mu$, where the bulk spectrum is expected to
be gapped due to Eq.~(\ref{eq:C3PB}). This transition will also be
understood as a type-II transition following the closing of a gap
between the edge bands at $E=0$.

\begin{figure}
	\begin{centering}
		\includegraphics[width=10.5 cm]{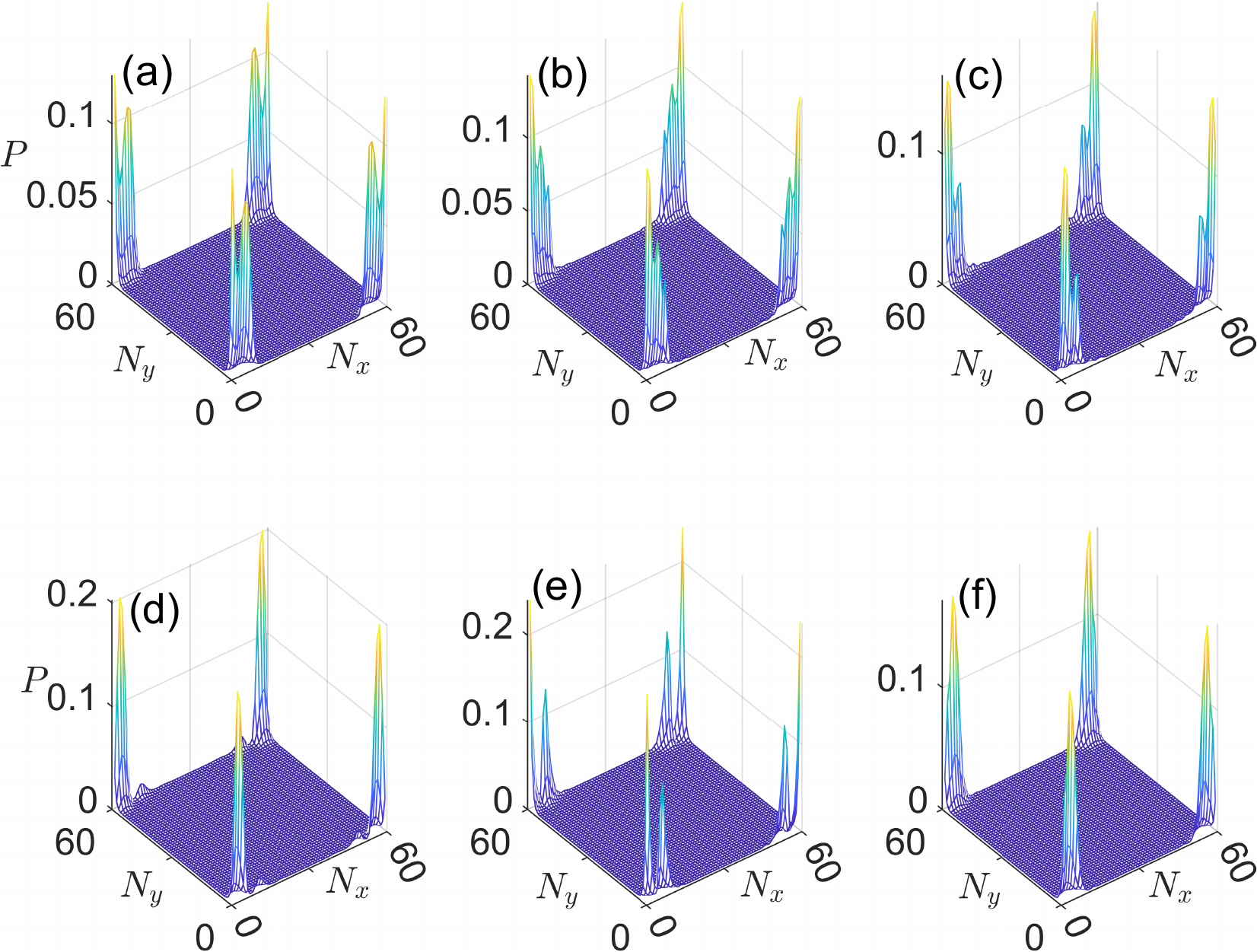}
		\par\end{centering}
	\caption{Probability distributions of Floquet corner modes in the Case 3 with
		quasienergies zero and $\pi$ in the panels (a)--(c) and the panels
		(d)--(f), respectively. Other system parameters are set as $(J,\mu,\Delta,\Delta_{1},\Delta_{2})=(3\pi,0.25\pi,0.5\pi,0.2\pi,0.4\pi)$.
		The number of cells along $x$ and $y$ directions are $N_{x}=N_{y}=60$.
		\label{fig:C3POBC}}
\end{figure}

In Fig.~\ref{fig:C3POBC}, we plot the probability distributions of
the first twelve and the last twelve of Floquet eigenmodes in 
Figs.~\ref{fig:C3EOBC}(b) and \ref{fig:C3EOBC}(c), respectively. It is clear
that they are indeed Floquet Majorana corner modes at the quasienergies
zero and $\pm\pi$. We could thus obtain many such quartets of Majorana
modes in our Floquet SOTSC system with $\Delta_{1}\neq0$. Furthermore,
their appearances follow transitions over gapless phases instead of
critical transition points of the system (except at $J=\mu$), which
are different from those happened in the Cases 1 and 2. The large
numbers of Majorana corner modes found here may also allow us to implement
Floquet quantum computing protocols in more general situations \cite{FHOTSC2}.

\begin{figure}
	\begin{centering}
		\includegraphics[width=10.5 cm]{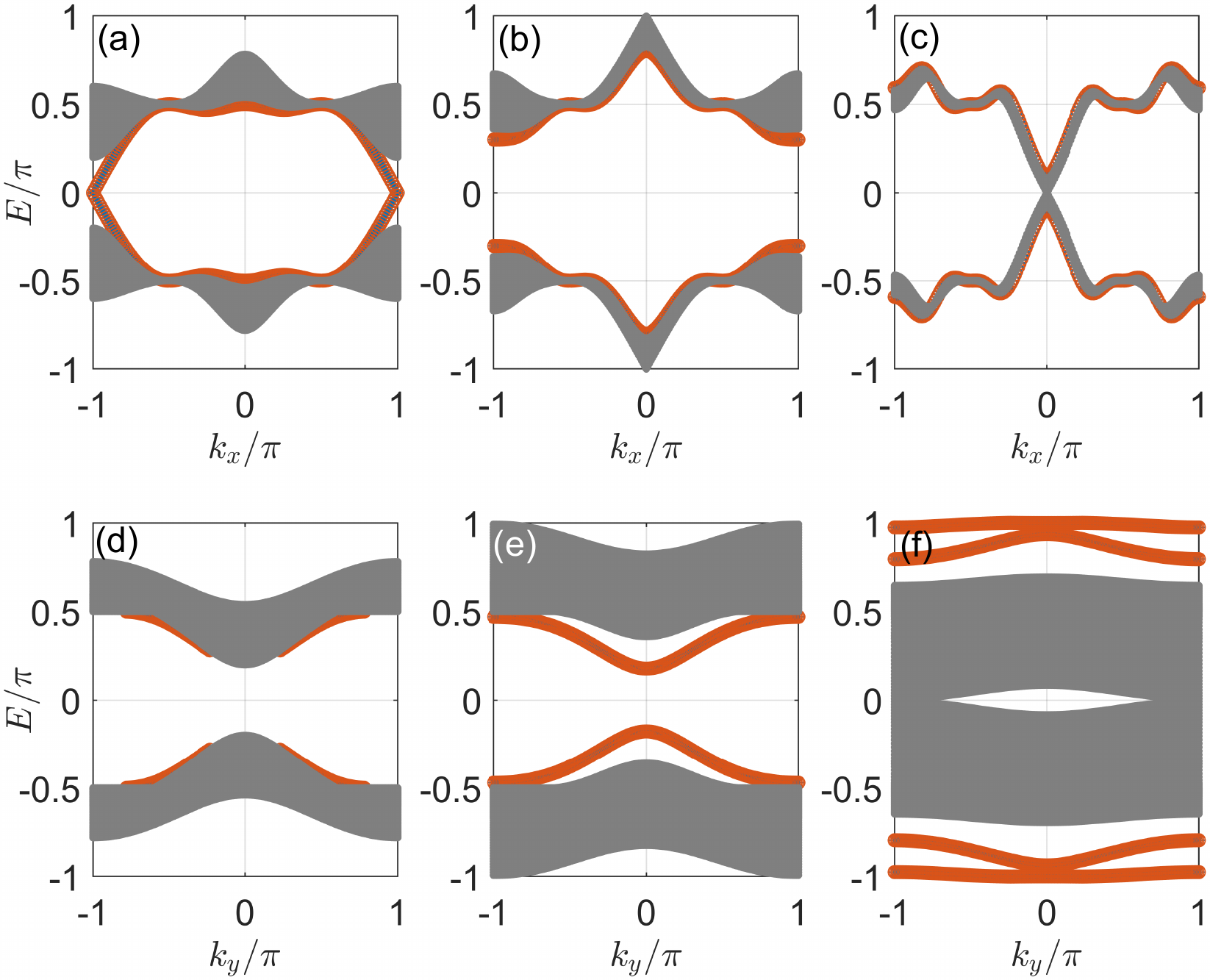}
		\par\end{centering}
	\caption{Floquet spectrum versus $k_{x}$ ($k_{y}$) in the Case 3 under PBCX,
		OBCY (OBCX, PBCX) in panels (a)--(c) {[}(d)--(f){]}. The gray dots,
		red circles and blue stars highlight the bulk states, states localized
		around the left edge and the right edge of the lattice. The value
		of hopping amplitude is set to $J=\mu$ for panels (a), (d), to the
		first bulk gap-closing point at $E=\pm\pi$ for panels (b), (e), and
		to the first bulk gap-closing point at $E=0$ for panels (c), (f).
		Other system parameters are set as $(\mu,\Delta,\Delta_{1},\Delta_{2})=(0.25\pi,0.5\pi,0.2\pi,0.4\pi)$.
		\label{fig:C3EMBC}}
\end{figure}

We are left to understand the transition at $J=\mu$. Similar to what
we have done in the last two subsections, we obtain the Floquet spectrum
of $\hat{U}(k_{x})$ and $\hat{U}(k_{y})$ in Eqs.~(\ref{eq:Ukx})
and (\ref{eq:Uky}) under the PBCX, OBCY and OBCX, PBCX, respectively,
and present them in Figs.~\ref{fig:C3EMBC}(a)--(c) and \ref{fig:C3EMBC}(d)--(f)
for $J=\mu$, $J=\sqrt{\pi^{2}-(\Delta_{1}+\Delta_{2})^{2}}-\mu$,
and $J=\sqrt{(2\pi)^{2}-(\Delta_{1}+\Delta_{2})^{2}}-\mu$. In the
latter two cases, the transitions happen through the touching of bulk
bands at $E=\pm\pi$ and $E=0$ with the quasimomentum $(k_{x},k_{y})=(0,\pi)$.
This coincides with the prediction of the bulk phase boundary in 
Eq.~(\ref{eq:C3PB}). They are thus conventional type-I topological phase
transition in the Floquet SOTSC system. Instead, we find that the
Floquet spectrum versus $k_{y}$ is gapped for $J=\mu$, while it
is gapless at $E=0$ through a touching of edge state bands at $k_{x}=\pi$.
Combining this observation with the emerging zero corner modes after
the first transition in Fig.~\ref{fig:C3EOBC}(a), we conclude that
the transition at $J=\mu$ is indeed a type-II topological phase transition
induced by an edge band touching, and it is not modified by the presence
of a finite intracell pairing strength $\Delta_{1}$ along the $y$
direction of the lattice. It deserves to be mentioned that the edge
bands in Fig.~\ref{fig:C3EMBC}(a) have vanishing net chiralities
even though they traverse the band gap, yielding Floquet bands with
zero Chern numbers.

Putting together, we find two types of topological phase transitions in
the Cases 1--3. With the increase of the hopping amplitude $J$,
more and more Floquet Majorana zero/$\pi$ corner modes emerge within
our system, and richer patterns of topological phases and transitions
appear in the cases with nonvanishing $\mu$ and $\Delta_{1}$. We
now arrive at the stage to discuss what will happen if a finite hopping
amplitude $J'$ along the $y$ direction is turned on. This will be
our task in the following subsection, in which we treat the effect
of $J'$ essentially as a perturbation.

\subsection{Case 4: General situation}\label{subsec:C4}

We finally discuss the general case with $J'\neq0$. The Floquet operator
of our model in momentum space now takes the form of Eq. (\ref{eq:Uk}), i.e.,

\begin{alignat}{1}
	U({\bf k})& = e^{-i\{(\mu+J\cos k_{x})\sigma_{0}\otimes\sigma_{z}+[\Delta_{2}\sin k_{y}\sigma_{x}+(\Delta_{1}-\Delta_{2}\cos k_{y})\sigma_{y}]\otimes\sigma_{x}+J'[(1+\cos k_{y})\sigma_{x}+\sin k_{y}\sigma_{y}]\otimes\sigma_{z}\}}\nonumber \\
	& \times e^{-i\Delta\sin k_{x}\sigma_{0}\otimes\sigma_{y}}.\label{eq:C4U}
\end{alignat}
In this case, a simple expression of the quasienergy bands cannot
be obtained from the Taylor expansion of $U({\bf k})$, as the tensor
products of Pauli matrices $\sigma_{x}\otimes\sigma_{z}$ and $\sigma_{y}\otimes\sigma_{z}$
are not anticommute with all other matrices. We thus devolve to numerical
calculations of the spectrum and states. Note that the gap functions
$F_{0}$ and $F_{\pi}$ for the Case 4 should be defined more generally as

\begin{equation}
	F_{0} \equiv \frac{1}{\pi}\min_{{\bf k}\in{\rm BZ}}\min_{j}|E_{j}({\bf k})|,\qquad
	F_{\pi} \equiv \frac{1}{\pi}\min_{{\bf k}\in{\rm BZ}}\min_{j}||E_{j}({\bf k})|-\pi|,\label{eq:F0P4}
\end{equation}
where $j=1,2,3,4$ labels all possible quasienergies of $U({\bf k})$
at each ${\bf k}$. 

\begin{figure}
	\begin{centering}
		\includegraphics[width=10.5 cm]{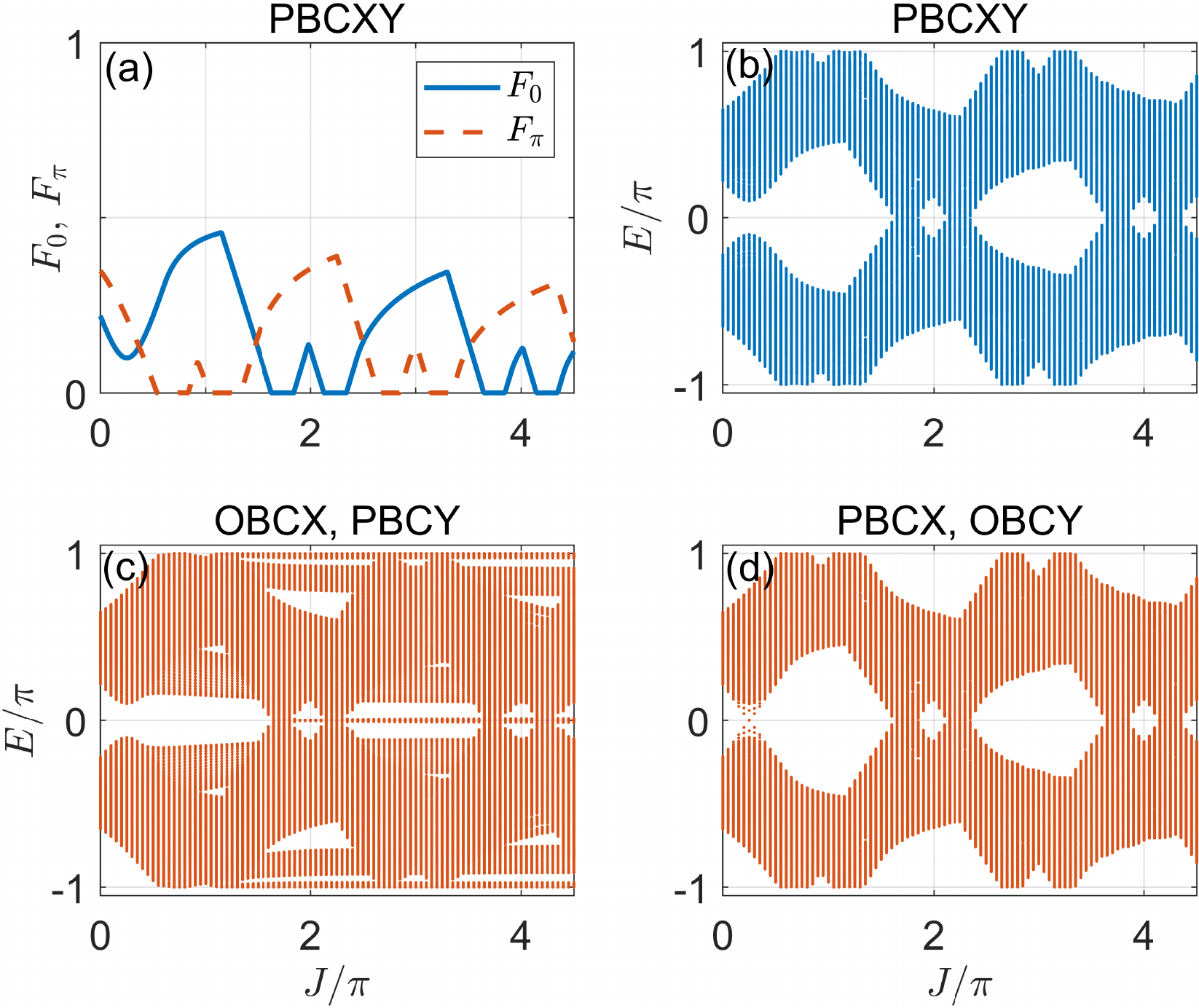}
		\par\end{centering}
	\caption{Gap functions and Floquet spectrum versus $J$ under different boundary
		conditions in the Case 4. (a) Gap functions under PBCXY. (b) Floquet
		spectrum under PBCXY. (c) Floquet spectrum under OBCX, PBCY. (d) Floquet
		spectrum under PBCX, OBCY. Other system parameters are set as $(\mu,\Delta,\Delta_{1},\Delta_{2},J')=(0.25\pi,0.5\pi,0.2\pi,0.4\pi,0.05\pi)$.
		The number of cells along $x$ and $y$ directions are $N_{x}=N_{y}=60$.
		\label{fig:C4GF}}
\end{figure}

In Fig.~\ref{fig:C4GF}(a), we show the gap functions versus $J$
in the Case 4 for a typical set of system parameters. We see that
with a nonvanishing but small $J'$, the general pattern of spectral
transitions in the Case 3 is preserved. That is, with the increase
of $J$, the system could enter a series of gapped Floquet phases
in the bulk (with $F_{0}\neq0$ and $F_{\pi}\neq0$), which are separated
by multiple gapless phases (with $F_{0}=0$ or $F_{\pi}=0$). The
main difference, caused by a finite $J'$, is that the regions of
gapless phases are broadened in the parameter space. This is also
confirmed by the Floquet spectrum versus $J$ plotted under different
boundary conditions in Figs.~\ref{fig:C4GF}(b)--(d). Therefore,
we expect the gapped Floquet SOTSC phases to be robust to certain
amounts of hoppings along the $y$ directions. Notably, the anomalous
transition at $J=\mu$, which is not related to the closing of a bulk
spectral gap is also observed in Fig.~\ref{fig:C4GF}(d), which implies
that this type-II topological transition remains intact when $J'\neq0$.

\begin{figure}
	\begin{centering}
		\includegraphics[width=10.5 cm]{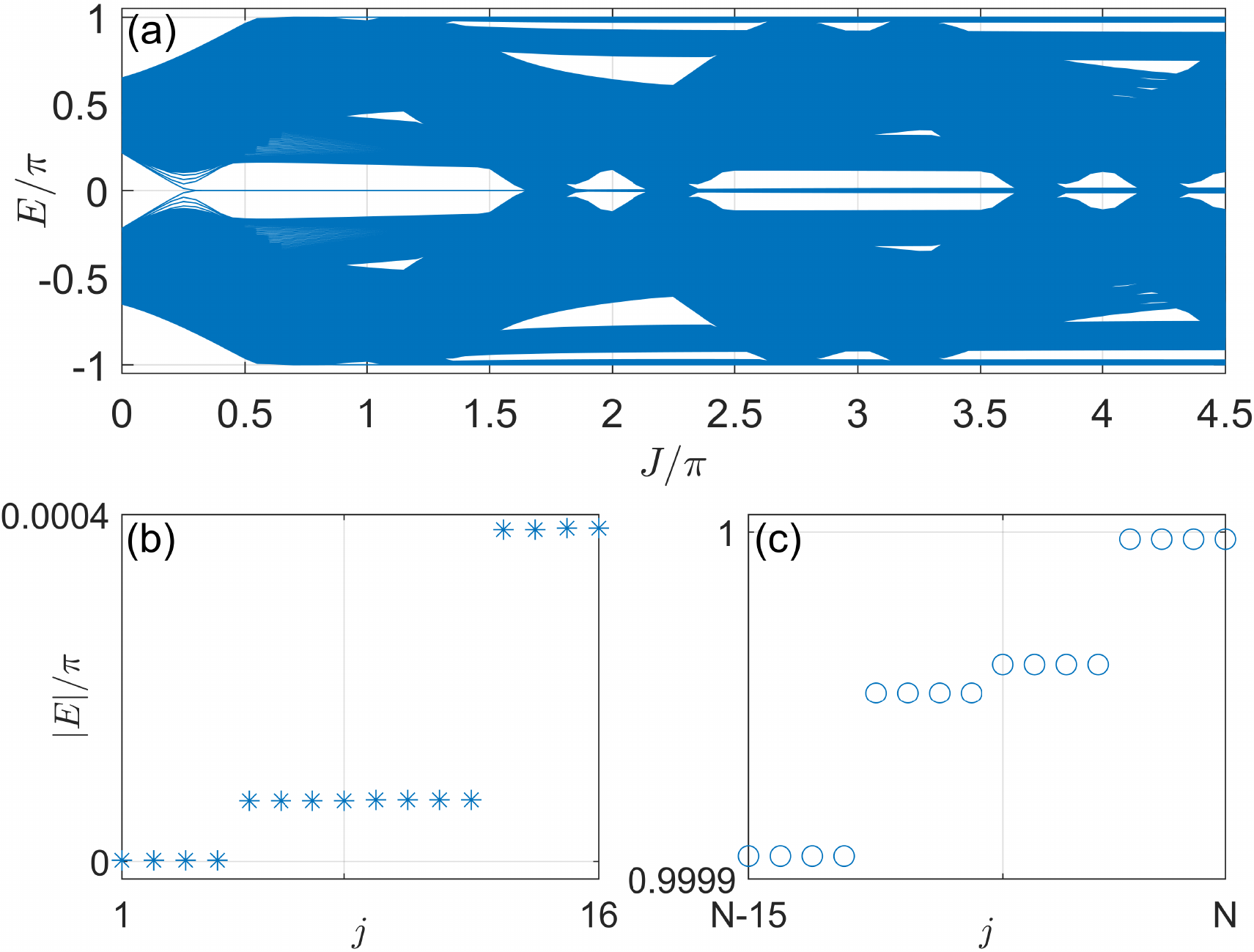}
		\par\end{centering}
	\caption{Floquet spectrum versus $J$ under OBCs in the Case 4. (a) Quasienergies
		at different $J$ under OBCXY. (b) and (c) show the absolute values
		of quasienergies of the first sixteen and last sixteen Floquet eigenstates
		indexed by $j$ at $J=3\pi$. Other system parameters are $(\mu,\Delta,\Delta_{1},\Delta_{2},J')=(0.25\pi,0.5\pi,0.2\pi,0.4\pi,0.05\pi)$.
		The number of cells along the $x$ and $y$ directions are $N_{x}=N_{y}=60$.
		The total number of Floquet eigenstates is $N=14400$. \label{fig:C4EOBC}}
\end{figure}

\begin{figure}
	\begin{centering}
		\includegraphics[width=10.5 cm]{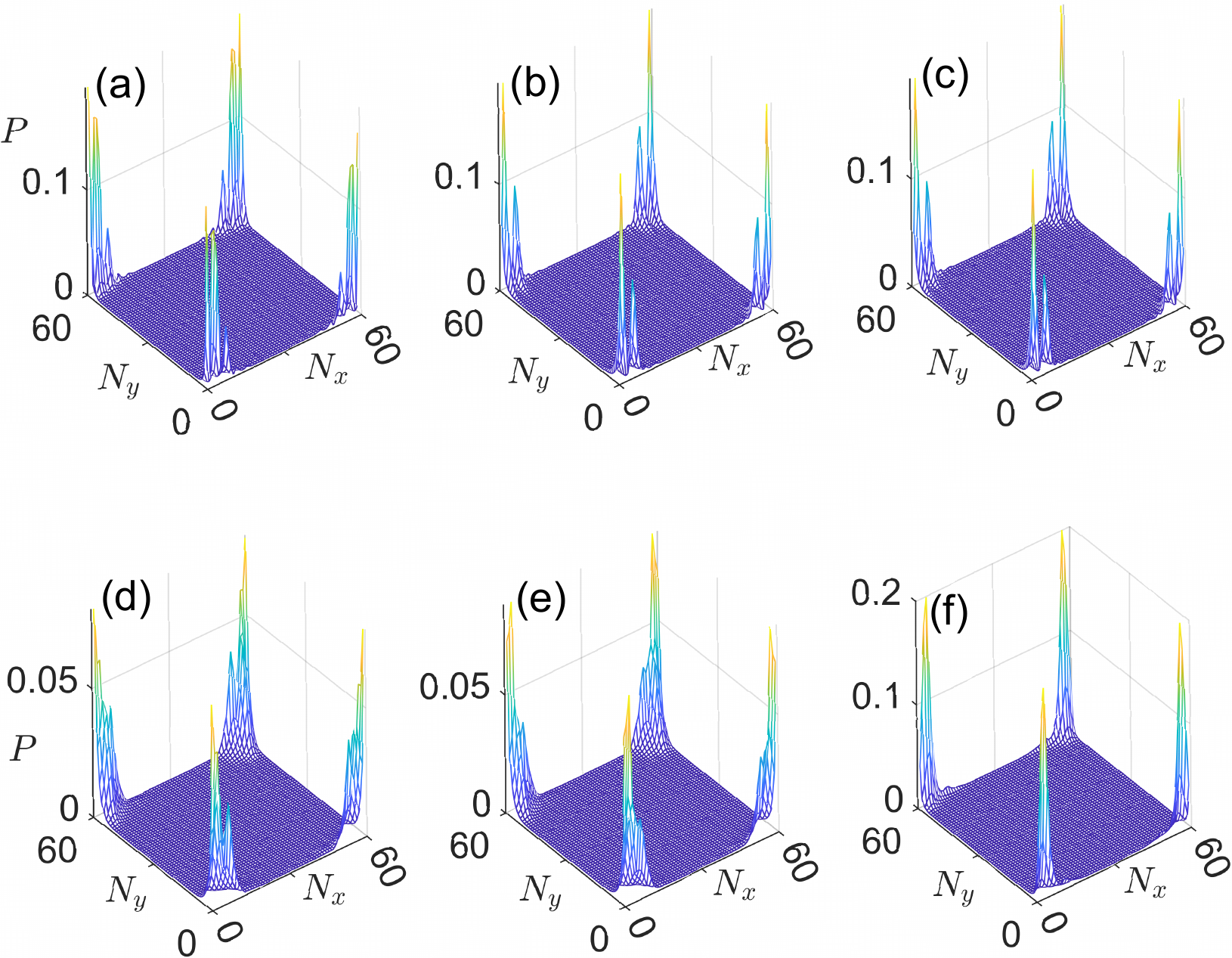}
		\par\end{centering}
	\caption{Probability distributions of Floquet corner modes in the Case 4 with quasienergies
		at or close to zero and $\pi$ in the panels (a)--(c) and the panels (d)--(f),
		respectively. Other system parameters are set as $(J,\mu,\Delta,\Delta_{1},\Delta_{2},J')=(3\pi,0.25\pi,0.5\pi,0.2\pi,0.4\pi,0.05\pi)$.
		The number of cells along $x$ and $y$ directions are $N_{x}=N_{y}=60$.
		\label{fig:C4POBC}}
\end{figure}

To further understand the fate of Floquet Majorana zero/$\pi$ corner
modes in the presence of a finite $J'$, we plot the spectrum under
OBCs along both the two spatial dimensions in Fig.~\ref{fig:C4EOBC}(a).
The results suggest that with the increase of $J$, there are still
eigenmodes with zero and $\pi$ quasienergies coming out of the gap-closing
transition at $J=\mu$ and the transitions through other gapless regions.
Meanwhile, in the gapped phases, we notice that the spectra become
broadened around $E=0$ and $E=\pm\pi$ compared with the Cases 1--3.
In Figs.~\ref{fig:C4EOBC}(b) and \ref{fig:C4EOBC}(c), we show the
absolute values of quasienergy for the first sixteen and the last
sixteen eigenstates in the spectrum at $J=3\pi$. In this case, we
only find four eigenmodes with the quasienergies $E=0$ and $\pm\pi$,
which means that not all the zero and $\pi$ Floquet Majorana corner
modes are robust to the perturbation introduced by $J'$. A possible
explanation of this observation is that with $J'\neq0$, the chiral
symmetry $\Gamma=\sigma_{z}\otimes\sigma_{x}$ of the Floquet system
is broken. Yet, the particle-hole symmetry ${\cal C}=\sigma_{0}\otimes\sigma_{x}{\cal K}$
is preserved as mentioned in Sec.~\ref{sec:Mod}, which enables the
system to possess at most four Floquet Majorana corner modes at the
quasienergies zero and $\pi$. However, since we do not expect a spectral
gap-closing between bulk or edge bands when $J'$ goes from zero up
to a small value ($0.05\pi$ here), the emerging eigenmodes away from
but close to $E=0$ and $\pm\pi$ in Figs.~\ref{fig:C4EOBC}(b) and
\ref{fig:C4EOBC}(c) should still have profiles localized around the
four corners of the lattice, since they have no chances to mix up
with other bulk or edge states under the perturbation.

In Figs.~\ref{fig:C4POBC}(a)--(c) and \ref{fig:C4POBC}(d)--(f),
we present the probability distributions of the first twelve and the
last twelve eigenstates in Figs.~\ref{fig:C4EOBC}(b) and \ref{fig:C4EOBC}(c),
respectively. Indeed, we find that all these modes are localized around
the four corners of the lattice, even though only eight of them are
Majorana zero and $\pi$ corner modes of the Floquet SOTSC phase {[}in
Figs.~\ref{fig:C4POBC}(a) and \ref{fig:C4POBC}(d){]}. The fourfold-degenerate
corner modes with $E\neq0,\pm\pi$ are not topologically protected,
as their quasienergies can vary with the change of $J'$. It remains
an interesting issue to explore whether these non-topological corner
modes could also find applications in Floquet quantum computing schemes.
In our calculations, we have also checked the effects of dimerized hopping amplitude and chemical potential on our results. These dimerization terms are introduced in the same way as the $\delta J$ and $\delta\mu_0$ used in the Eq.~(1) of Ref.~\cite{FHOTSC2}. Our numerical results suggest that the main observations about Floquet spectrum and corner modes presented in this subsection hold when these dimerization terms are introduced to our model as small perturbations.

\section{Discussion and Conclusion}\label{sec:Sum}

In this work, we explored the generation of many Majorana zero/$\pi$
corner modes and multiple phase transitions in Floquet second-order
topological superconductors. By applying time-periodic kicking to
the pairing amplitude of a generic 2D $p_{x}+ip_{y}$ superconductor
with dimerized superconducting pairing, we obtain rich Floquet SOTSC
phases with arbitrarily many Majorana corner modes at zero and $\pi$
quasieneriges in principle in the thermodynamic limit. Moreover, two
different Floquet SOTSC phases are found to be separated by either
a type-I topological phase transition with the closing/reopening of
a bulk spectral gap, a type-II topological phase transition with the
closing/reopening of an edge band gap, or a third phase with gapless
quasienergy spectrum. The multiple quartets of Floquet Majorana zero/$\pi$
corner modes are found to be protected by the chiral and particle-hole
symmetries of the system. When the chiral symmetry is lifted by a
perturbation that couples different sites along the $y$-direction,
four out of the many Majorana zero and $\pi$ corner modes could still
survive, which represent the most stable higher-order topological
modes in the system. Our work thus extends the present understanding
of topological phases and transitions in Floquet SOTSC setups. It
further provides an efficient means to generate many Majorana corner
modes, which may give us more choices and freedoms to design topological
quantum computing schemes.

From the perspective of bulk-corner correspondence, there should exist a pair of integer-quantized topological invariants that can characterize the two types of Floquet Majorana corner modes at zero and $\pi$ quasienergies in our system, similar to what has been identified before for Floquet second-order topological insulators \cite{FHOTP1}. In practice, due to the complicated patterns of phase transitions and Floquet corner modes in our model, the topological winding numbers introduced in Ref.~\cite{FHOTP1} could not fully characterize all the Majorana corner modes and bulk-corner correspondence in our system. In future work, it remains an interesting issue to explore suitable topological invariants that could predict the numbers of Majorana corner modes in different Floquet SOTSC phases found here.

In our Floquet model, the periodic driving is introduced via a delta-kicked superconducting pairing amplitude. We expect similar results as those reported in Sec.~\ref{sec:Res} if the delta-kick is replaced by a piecewise quench applied successively to $\Delta$ and the other terms of ${\hat H}$ within each driving period. With a sinusoidal driving field, much fewer Floquet zero and $\pi$ Majorana corner modes and more restricted patterns of topological phase transitions are expected in our system, similar to what has been identified before in Ref.~\cite{FHOTSC2}. Meanwhile, in the presence of two sinusoidal drivings with different frequencies, there are still opportunities to find rich Floquet phases with many topological boundary states and multiple topological transitions \cite{LiuPump2022}. It is thus interesting to consider the Floquet engineering of SOTSC phases by multi-frequency harmonic driving fields in future studies.

In our system, we have numerically checked that when the time-periodic delta kicking is applied to the hopping amplitude $J$ instead of the pairing amplitude $\Delta$, similar results could be obtained as those reported in Sec.~\ref{sec:Res}. That is, we can now find many Floquet Majorana zero and $\pi$ corner modes together with multiple topological phase transitions with the increase of $\Delta$ when $J$ is kicked. Interestingly, substantial differences were observed with the periodic delta kicking applied to the pairing versus hopping amplitudes in the chaotic dynamics of interacting Floquet models~\cite{RoyPRE2020}. Therefore, the signature of quantum chaos in many-body Floquet SOTSC systems under different kicking protocols deserves to be revealed in future studies.

The effects of disorder and boundary confinements form major challenges in the experimental study of Majorana bound states in topological superconductors \cite{Dsod1,Dsod2}. In the present work, we focus on the theoretical possibility of generating many Majorana zero/$\pi$ corner modes and multiple topological phase transitions with the help of Floquet driving fields. For completeness, we have also numerically checked the effects of disorder on Majorana corner modes for the different cases considered in Sec.~\ref{sec:Res}. We found that for the Cases 1--3, the Floquet Majorana corner modes are robust to disorder in the forms of $\sum_{m,n}\delta\mu_{m,n}(\hat{a}_{m,n}^{\dagger}\hat{a}_{m,n}+\hat{b}_{m,n}^{\dagger}\hat{b}_{m,n})$, $\sum_{m,n}\delta J_m(\hat{a}_{m,n}^{\dagger}\hat{a}_{m+1,n}+\hat{b}_{m,n}^{\dagger}\hat{b}_{m+1,n}+{\rm H.c.})$, $\sum_{m,n}\delta\Delta_m(\hat{a}_{m,n}\hat{a}_{m+1,n}+\hat{b}_{m,n}\hat{b}_{m+1,n}+{\rm H.c.})$ and $\sum_{m,n}(i\delta\Delta_{1,n}\hat{a}_{m,n}\hat{b}_{m,n}+i\delta\Delta_{2,n}\hat{b}_{m,n}\hat{a}_{m,n+1}+{\rm H.c.})$. Here $\delta\mu_{m,n}$, $\delta J_m$, $\delta\Delta_m$, $\delta\Delta_{1,n}$ and $\delta\Delta_{2,n}$ vary randomly in space over different sites and bonds. The values for each of them are taken separately from a uniform distribution with the range $[-W,W]$ (we take $W=1/20$ in our numerical calculations). Note that these forms of disorder do not break the chiral symmetry of the model. The many Floquet Majorana corner modes in our system are found to be well localized around the corners so long as they are well separated in quasienergies from the other bulk or edge states in the presence of disorder. In the Case 4, the four Majorana zero/$\pi$ corner modes are further found to be robust to weak disorder in the form of $\sum_{m,n}\delta J'_n(\hat{a}_{m,n}^{\dagger}\hat{b}_{m,n}+\hat{b}_{m,n}^{\dagger}\hat{a}_{m,n+1}+{\rm H.c.})$, where $\delta J'_n$ varies over bonds along the $y$-direction and taking values randomly in a uniform distribution with the range $[-W,W]$. Other forms of disorder and boundary confinements may also change the fate of Majorana corner modes in our system, which deserve to be investigated more thoroughly in a potential future work.

Finally, it would be interesting
to explore the engineering of Floquet SOTSC phases in other symmetry
classes, in higher spatial dimensions and with many-body interactions.
The application of the multiple Floquet zero/$\pi$ Majorana corner
modes found here to different quantum computing protocols also deserves
to be investigated in detail.



\vspace{6pt} 




\section*{Acknowledgement}
This research was funded by the National Natural Science Foundation of China (Grant No.~11905211, 12275260), the Young Talents Project at Ocean University of China (Grant No.~861801013196), and the Applied Research Project of Postdoctoral Fellows in Qingdao (Grant No.~861905040009).

\end{document}